\documentclass[%
 reprint,
 amsmath,amssymb,
 aps,
]{revtex4-2}
\usepackage{graphicx}
\usepackage{dcolumn}
\usepackage{bm}
\usepackage{color}
\usepackage[normalem]{ulem}
\usepackage[
  colorlinks=true,
  linkcolor=black,
  citecolor=black,
  urlcolor=black
]{hyperref}

\newcommand{\removed}[1]{}
\definecolor{ashgrey}{rgb}{0.7, 0.75, 0.71}

\begin{document}
\preprint{APS/123-QED}	
	\title{Intrinsic (non)-Gilbert damping in magnetic insulators calculated from a minimal model and \textit{ab initio} spin Hamiltonians.}

	\author{Andrei Shumilin}
    \email{andrei.shumilin@uv.es}
    \author{ Diego López-Alcalá,  	
     Nassima Benchtaber, Alberto M. Ruiz}
    \author{José J. Baldoví}
     \email{J.Jaime.Baldovi@uv.es}
\affiliation{Instituto de Ciencia Molecular, Universitat de València, 46980 Paterna, Spain}

\begin{abstract}
We present an analytically solvable minimal model for the relaxation of low-frequency magnons in magnetic insulators arising from magnon–phonon and magnon–magnon interactions. The model establishes a direct connection between microscopic relaxation processes and Gilbert damping, and reveals how magnon decay evolves from bulk systems to the monolayer limit. We find that magnon–phonon coupling produces Gilbert damping of comparable magnitude in three- and two-dimensional magnets, with qualitative differences between flexural phonons in free-standing monolayers and three-dimensional phonons in substrate-supported layers. By contrast, non-Gilbert damping due to four-magnon scattering is strongly enhanced in two dimensions, where it becomes independent of spin–orbit coupling. To benchmark the model against real materials, we introduce a numerical approach for computing magnon damping from ab initio–derived spin Hamiltonians. We demonstrate that the central conclusions of the model remain valid for magnons in bulk YIG and several van der Waals magnetic insulators: $\rm CrSBr$, $\rm CrI_3$ and $\rm CrPS_4$.	
\end{abstract}

\maketitle
	
\section{Introduction}

Magnonics is a rapidly advancing branch of magnetoelectronics and spintronics that focuses on the generation, control, and detection of magnons—the quanta of spin waves~\cite{map21,map24}. It offers a versatile platform for investigating fundamental spin phenomena and enables a broad range of device concepts, including low-power information processing~\cite{mapQ} and artificial neural networks~\cite{neu2021}. Early developments in magnonics relied primarily on relatively thick magnetic films combined with macroscopic magnon antennas~\cite{antena}. However, the ongoing trend toward miniaturization has opened new possibilities, including quasi-two-dimensional (2D) van der Waals (vdW) magnets, which retain magnetic order down to the monolayer limit~\cite{monolayer}.

Despite this progress, a major challenge common to all magnonic technologies is the relaxation (or damping) of spin waves. This relaxation is most commonly incorporated into the Landau--Lifshitz--Gilbert (LLG) equation---the standard framework for spin-wave simulations---through the Gilbert damping parameter $\alpha$~\cite{gilbert,Aharoni1996}. Although several strategies for magnon amplification have been proposed~\cite{PumpingPirro,Pumping2,torque1}, low damping remains a key requirement for practical magnonic materials. This has made the insulating ferrimagnet yttrium iron garnet (YIG), with its exceptionally low damping $\alpha<10^{-4}$, the prototypical magnonic material. The search for new low-damping magnetic insulators therefore continues, with several promising candidates proposed \cite{AdvCohMag}, including insulating vdW layered magnets such as ${\rm CrSBr}$ and ${\rm CrPS_4}$ \cite{CrSBr0,longCrPS4}. Magnetic insulators are particularly attractive because the absence of itinerant electrons eliminates magnon--electron damping. Moreover, since Gilbert damping is often associated with spin--orbit coupling (SOC) \cite{Kambersk,Moodera,Yudin}, materials with weak SOC seem to be especially promising.

Experimentally, the damping parameter $\alpha$ is most commonly extracted from ferromagnetic resonance (FMR) measurements via the magnetic-field dependence of the linewidth. The FMR linewidth typically consists of two main contributions: an inhomogeneous component related to material nonuniformity and independent of frequency, and a homogeneous component associated with magnon decay~\cite{Gurevich}. If the decay is governed by Gilbert damping, the homogeneous linewidth is proportional to the FMR frequency, enabling its separation from the inhomogeneous contribution. However, several experimental studies have reported deviations from this so-called Gilbert behavior~\cite{NiFe,LSMO,LSMO2}, including linewidths that decrease with increasing FMR frequency~\cite{YIGlove}. One well-known source of such non-Gilbert behavior is two-magnon scattering induced by impurities or surface roughness~\cite{TMSultra,TMS2004,TMS2024}. Beyond FMR, spin-wave decay can also be probed directly using time-resolved techniques, which allow the relaxation rate to be determined independently of its frequency dependence \cite{tL1,tNiFeGd,tFeGas}.

In many experiments, the measured damping is dominated by extrinsic effects arising from disorder and impurities~\cite{impNiFe,impYIG}. Intrinsic damping, however, sets the fundamental lower bound in a perfect crystal. Its three main microscopic mechanisms are magnon--electron, magnon--phonon, and magnon--magnon scattering. Magnon--electron scattering has been studied most extensively and can be evaluated using several \textit{ab initio} approaches \cite{Moodera,abin2}, typically yielding $\alpha\sim10^{-3}$~\cite{abinLayered,FeGaTe}. However, this mechanism is absent in insulating magnets, motivating the search for the remaining intrinsic damping channels. Magnon–phonon interactions have been theoretically investigated in several works~\cite{atomistic,CrI3,nonlinear,goodYIG19,goodYIG2014}, with some of these studies enabling the extraction of phonon-induced contributions to the Gilbert damping in YIG~\cite{goodYIG19,goodYIG2014}. Magnon–magnon scattering has likewise been examined~\cite{magmag,magmagCrI3}, often with particular emphasis on antiferromagnetic systems~\cite{MMafm76,MMafm13}. However, these studies typically focus on specific materials and frequently address magnons at large wave vectors, whereas Gilbert damping is defined in the low-wavevector limit. Consequently, a unified understanding of which intrinsic damping mechanisms dominate in a given insulating material—and how their relative importance evolves from bulk systems to the monolayer limit—remains lacking.

In this Letter, we present a systematic study of intrinsic damping mechanisms in magnetic insulators, with particular emphasis on the differences between two- and three-dimensional (2D and 3D) materials and comparison with LLG. We employ the Boltzmann formalism, which is widely used in the theory of magnon relaxation \cite{goodYIG19,goodYIG2014,magmagCrI3}. It relies on the assumption of weak magnon spectral broadening, such that the magnon decay rate remains much smaller than the magnon frequency \cite{LandauKin, SmithJensen1989, Liboff}. In the present case, this requirement reduces to $\alpha \ll 1$, which is precisely the regime relevant for low-damping materials suitable for magnonics.

\section{general theory and the minimal toy model}

We consider a magnetic material described with a generalized  spin Hamiltonian 
\begin{equation}\label{SHgen}
H = -\sum_{i,j} {\bf S}_i \widehat{J}_{ij} {\bf S}_j - \sum_i \mu_b g_i {\bf S}_i {\bf B} - \sum_{i,j;k} \mathbf{S}_i \frac{\partial \widehat{J}_{ij}  }{\partial {\bf r}_k} {\bf S}_j {\bf u}_k.
\end{equation}
Here ${\bf S}_i$ denotes the vector of spin operators associated with a magnetic atom $i$, and $g_i$
 is the corresponding Lande g-factor. The tensor
$\widehat{J}_{ij}$ describes the exchange interaction between magnetic atoms $i$ and $j$. While in the non-relativistic limit,
$\widehat{J}_{ij}$ reduces to a scalar, in the general case it becomes a full $3\times 3$ matrix in Cartesian coordinates~\cite{spinW}. In this general form, $\widehat{J}_{ij}$ can account both for the exchange interaction, including its anisotropic part, and magnetic dipole–dipole interactions. $\mu_b$ is the Bohr magneton and ${\bf B}$ denotes the external magnetic field. To account for the interaction with phonons, we consider the modulation of the exchange tensor 
$\widehat{J}_{ij}$ due to the atomic displacement ${\bf u}_k$, where the index $k$ runs over all the atoms of the system. $\partial \widehat{J}_{ij}/\partial {\bf r}_k$ is the derivative of the exchange tensor $\widehat{J}_{ij}$ over the position of atom $k$. Close to the ground state, Eq.~(\ref{SHgen}) can be mapped onto a quasiparticle Hamiltonian using the Holstein–Primakoff transformation~\cite{Holstein} and the second-quantization of the displacements ${\bf u}_k$. 

\begin{figure}
 	\centering
 	\includegraphics[width=3.4in]{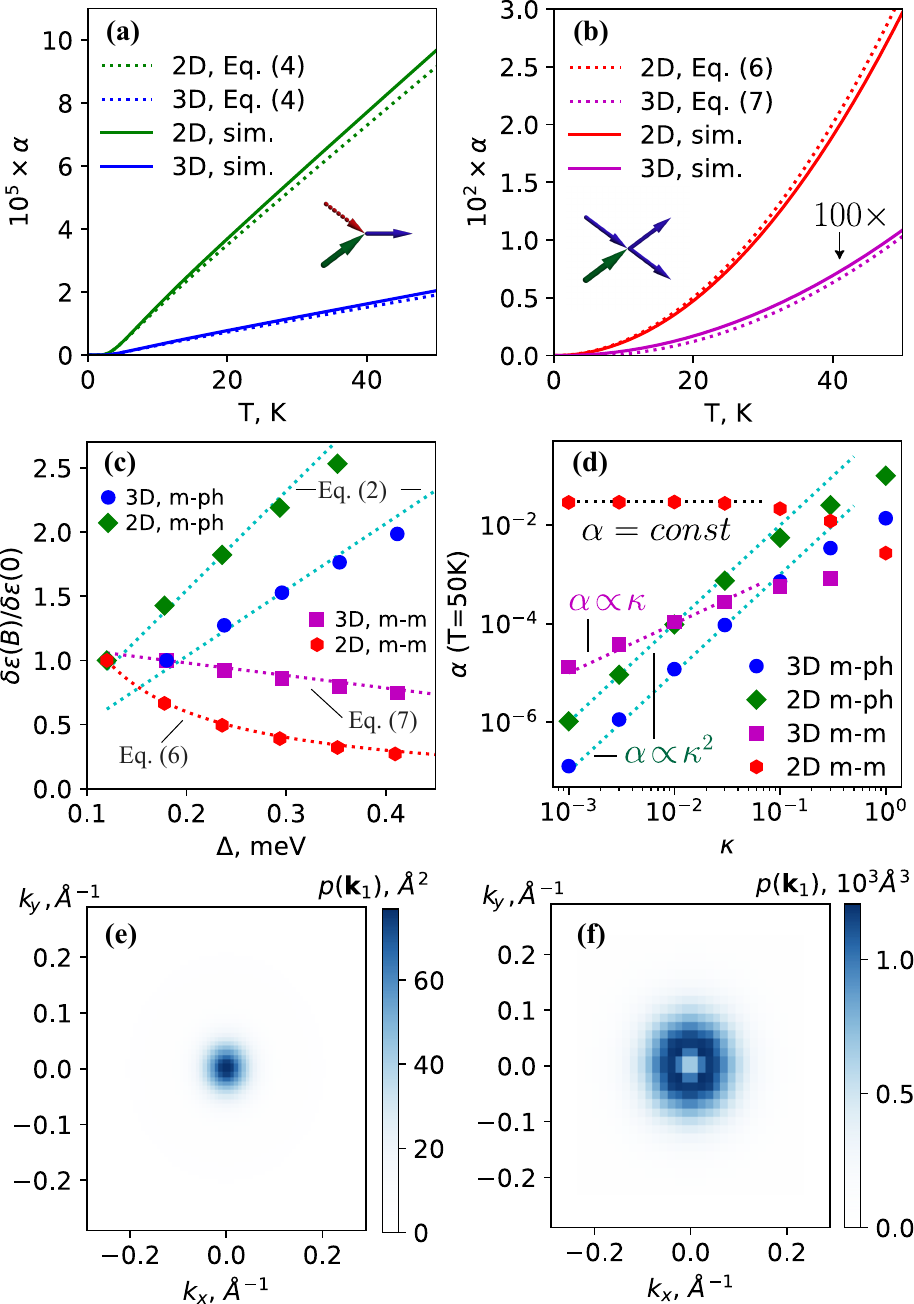}
 	\caption{ \label{fig:model}
 		Magnon damping in the minimal toy model. (a) Temperature dependence of the damping due to the magnon-phonon interaction. (b) Temperature dependence of the damping due to the magnon-magnon interaction. (c) Magnon linewidth dependence on the energy compared with the Gilbert law (cyan dotted lines) and with the non-Gilbert behavior derived from Eqs.~(\ref{mag2D}) and (\ref{mag3D}), red dotted and magenta dotted lines, respectively. (d) Dependence of the damping parameter $\alpha$ in zero magnetic field on the anisotropy parameter $\kappa$. (e,f) distribution of the magnons responsible for the magnon-magnon induced damping in 2D and 3D models, respectively.  } 
\end{figure}

We consider an initial excitation of a classical  spin-wave corresponding to a macroscopically large occupation number $N_0$ of ${\bf k} \rightarrow 0$ acoustic magnons. The occupation numbers of all other quasiparticles are thermal and follow the Bose–Einstein distribution. The relaxation of the initial spin-wave is described by the relaxation rate $\gamma_0$ proportional to the homogeneous linewidth in FMR experiments $\delta \varepsilon = \hbar \gamma_0$. An equivalent treatment of the same physical problem within LLG formalism yields the relation
\begin{equation} \label{Gilb}
    \gamma_0 = \delta\varepsilon/\hbar = \alpha \Delta/\hbar
\end{equation}
where $\Delta$ denotes the energy of the ${\bf k}=0$ acoustic magnons (the magnon gap), and 
$\alpha$ is the dimensionless Gilbert damping parameter.

To provide a unified description of both Gilbert and non-Gilbert magnon relaxation, we define the dimensionless damping parameter as $\alpha = \hbar \gamma_0/\Delta$
which can be evaluated at zero or finite external magnetic field. Within this framework, we distinguish between Gilbert damping, for which $\alpha$ is field-independent at least in small magnetic fields, and non-Gilbert damping corresponding to the relation (\ref{Gilb}) being broken at any values of $B$.

We begin our analysis from a minimal toy model, which consists of a square lattice in 2D or a cubic lattice in 3D, with lattice constant $a$, magnetic atoms of spin $S$, and atomic mass $m$.
We assume the ferromagnetic exchange interactions between nearest-neighbors:
\begin{equation}
{\bf S}_i \widehat{J}_{ij} {\bf S}_j = 
J\left({\bf S}_i \cdot {\bf S}_j + \kappa S_i^{(z)} S_j^{(z)}\right).
\end{equation}
Here $J$ denotes the  Heisenberg exchange coupling and $\kappa J$ represents a small uniaxial anisotropy along the $z$ axis. The anisotropy observed in typical magnetic materials corresponds to $\kappa \sim 0.04$~\cite{CrI3model} or smaller.~\cite{CrSBr0,CrPS4model} The lattice dynamics is modeled by three acoustic phonon branches with identical sound velocity $c$. The exchange interaction depends solely on the distance between neighboring atoms, and its modulation due to atomic displacements is characterized by a derivative $dJ/dr = J'$. The anisotropy parameter $\kappa$ is independent of atomic displacements. 

Within this model, both damping mechanisms allow analytical treatment when temperature $T$ is large compared to $\Delta$. The magnon-phonon interaction contributes to the damping  via the processes when the ${\bf k}=0$ acoustic magnon absorbs a phonon with a wavevector ${\bf q}$. The resulting expression for the Gilbert damping is~\cite{SI}
\begin{multline}\label{phon23D}
\alpha_{m-ph} =  2D
\frac{ k_1^{D-2}a^D (\kappa S J')^2}{(2\pi)^{D-1} \Delta m c^2  } \\ \times
\bigl[M(k_1) - N(k_1)\bigr]  {\cal J}_{D} (k_1 a) .
\end{multline}
Here $D$ is the dimensionality of the system. $k_1 = \hbar c/ J Sa^2$ is the phonon wave vector that satisfies the energy and momentum conservation laws.  The quantities $M(k_1)$ and $N(k_1)$ denote the Bose–Einstein occupation numbers of phonons and magnons, respectively. 
The functions ${\cal J}_{2}(x)$ and $ {\cal J}_{3}(x)$ are given by the expressions:
\begin{equation} \label{J2J3}
\begin{array}{l}
{\cal J}_{2}(x) = \int_0^{2\pi} \sin^2(x\cos(\phi))d\phi \approx \pi x^2 \\ 
{\cal J}_{3}(x) = (2\pi/x)(x - \cos(x)\sin(x)) \approx 4\pi x^2/3
\end{array}
\end{equation}
The r.h.s. of Eqs.~(\ref{J2J3}) corresponds to the condition $x \ll 1$. 

Fig.~\ref{fig:model}(a) shows the temperature dependence of the magnon damping calculated with Eq.~(\ref{phon23D}) compared with the numerical simulations of 2D and 3D model. The model parameters are:  $J= 2 \, \rm{meV}$,    
$\kappa = 0.01$, $a=3\,{\rm \AA}$, $c=910\,{\rm m/s}$, $m = 1\,\rm{a.e.m.}$; $J'=-J/a$. The damping is suppressed at $T < \hbar ck_1$ and grows linearly with temperature at larger values of $T$. It is somewhat larger in 2D case reaching $\sim 10^{-4}$ at a temperature of $50\,{\rm K}$, however, the results for both the 2D and 3D cases are comparable in order of magnitude.

\begin{figure*}
 	\centering
 	\includegraphics[width=6.5in]{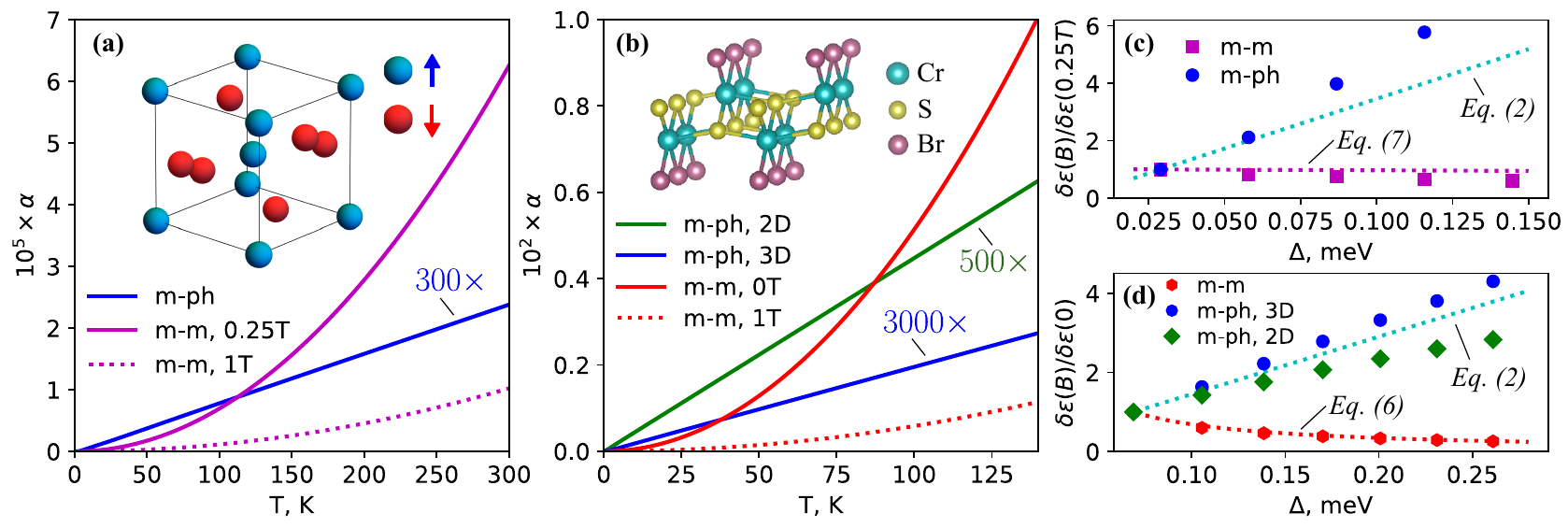}
 	\caption{ \label{fig:DFT}
        Magnon damping calculated from \textit{ab initio} spin Hamiltonians.
        (a) Temperature-dependent damping in bulk YIG arising from magnon–magnon (m–m) interactions at 
        $B=0.25\, {\rm T}$  and $B=1\, {\rm T}$, and from magnon–phonon (m–ph) interactions.
        Inset shows the positions of spin-up and spin-down Fe atoms in an 1/8 of YIG cubic cell.
        (b) Temperature-dependent damping in a CrSBr monolayer due to magnon–magnon (m–m) interactions at 
        $B=0$ and $B=1\, {\rm T}$, and due to magnon–phonon (m–ph) interactions with both 2D and 3D phonons. The inset illustrates the monolayer crystal structure.
        (c) FMR linewidth $\delta\varepsilon$ in YIG, normalized to its value at 
        $B=0.25\, {\rm T}$.  The m–ph and m–m contributions are compared with the Gilbert damping and with Eq.~(\ref{mag3D}), respectively. (d) FMR linewidth $\delta\varepsilon$ in a CrSBr monolayer, normalized to its value at $B=0$. The m–ph and m–m contributions are compared with the Gilbert damping and with Eq.~(\ref{mag2D}), respectively. Results for both 2D and 3D phonons are shown, as indicated in the legend. } 
\end{figure*}

The magnon-magnon interaction contributes to the damping via the magnon-conserving (MC) 4-magnon process, illustrated in Fig.~\ref{fig:end1} in Appendix~\ref{bol-Gen}.
The corresponding contributions to $\alpha$ are~\cite{SI}:
\begin{equation}\label{mag2D}
\alpha_{m-m}^{(2D)} = 
\frac{1.35 T^2 \Delta_0^2}{(2 \pi J S^2 )^2\Delta^2},
\end{equation}
\begin{equation}\label{mag3D}
\alpha_{m-m}^{(3D)} = \frac{\Delta_0^2 T^2 }{(2\pi J S^2  )^3\Delta}\left(0.82 - 2.94 \frac{\Delta}{T} \right).
\end{equation}
Here $\Delta_0 = 2D\kappa J S$ is the zero-field magnon gap with $\Delta = \Delta_0 + \mu_b g B$. Eqs.~(\ref{mag2D},\ref{mag3D}) include the quadratic temperature dependence of $\alpha$ as shown in Fig.~\ref{fig:model}(b). The damping is very strong in 2D case reaching $\sim 0.03$ at  $T = 50\,{\rm K}$. Nevertheless, it is $\sim 300$ times weaker in the 3D case.

Both in 2D and 3D model the magnon-phonon contribution to the damping follows the Gilbert law~(\ref{Gilb}) as shown in Fig.~\ref{fig:model}(c), where we present the FMR linewidth calculated in external magnetic fields up to $1.25\,{\rm T}$ and normalized to its value at zero magnetic field. However, the magnon-magnon contribution for both the dimensionalities is non-Gilbert due to the explicit dependence of $\alpha_{m-m}^{(2D)}$ and $\alpha_{m-m}^{(3D)}$, described by Eqs.~(\ref{mag2D}) and (\ref{mag3D}), respectively, on $\Delta$. In both the cases, the linewidth $\delta \varepsilon$ decreases with $\Delta$ as shown in Fig.~\ref{fig:model}(c). In 2D it follows the power law $\delta \varepsilon \propto \Delta^{-1}$ while in 3D this decrease is somewhat weaker. 

The strong damping due to the magnon-magnon mechanism in the 2D model is related to $\alpha_{m-m}^{(2D)}$ described by Eq.~(\ref{mag2D}) being independent of the anisotropy parameter $\kappa$ at zero external field. To illustrate this we show in Fig.~\ref{fig:model}(d) the damping calculated from both the mechanisms in 2D and 3D models with varying $\kappa$ and other parameters corresponding to  Figs.~\ref{fig:model}(a-c). For relatively large values of $\kappa$ all the mechanisms lead to comparable damping parameters. Nevertheless, at small $\kappa$ the magnon-phonon-induced $\alpha$ decreases as $\alpha\propto \kappa^2$, magnon-magnon mechanism in 3D yields a slower decrease $\alpha \propto\kappa$ and in 2D it leads to $\alpha \approx const$. 

As shown in Appendix~\ref{bol-Gen}, the non-Gilbert damping derived from the Boltzmann equation is governed by low-energy quasiparticles participating in scattering processes. Figures~\ref{fig:model}(e) and (f) display the  relative contributions $p({\bf k}_1)$  of magnons with different wavevectors ${\bf k}_1$ to the magnon–magnon damping mechanism in 2D and 3D toy models, respectively. In both cases, the damping is dominated by low-$|{\bf k}_1|$ magnons with energies close to the magnon gap, whose strong field dependence leads to non-Gilbert behavior. However, owing to the different magnon statistics, the contribution vanishes exactly at ${\bf k}_1 = 0$ in the 3D case [Fig.~\ref{fig:model}(f)], while no such suppression occurs in 2D [Fig.~\ref{fig:model}(e)], rendering the non-Gilbert behavior less pronounced in three dimensions.

\section{Modeling of real materials}

To demonstrate the relevance of our results to real materials, we simulate the damping in several magnonic systems using parameters obtained from \textit{ab initio} calculations, as described in Appendix~\ref{numDet} and in the Supporting Information (SI)~\cite{SI}. We begin with bulk YIG and a monolayer of ${\rm CrSBr}$ to highlight the contrast between three- and two-dimensional systems.

YIG is a ferrimagnetic insulating oxide crystallizing in the cubic centrosymmetric space group ${\rm Ia\overline{3}d}$. Its magnetic order arises from two antiferromagnetically coupled ${\rm Fe^{3+}}$ sublattices (octahedral and tetrahedral sites) with unequal occupation (24 and 16 Fe ions per cubic unit cell), leading to a finite net magnetization despite the antiparallel spin alignment (see inset in Fig.~\ref{fig:DFT}(a)).
The temperature dependence of magnon–magnon and magnon–phonon damping in bulk YIG is shown in Fig.~\ref{fig:DFT}(a). The results are consistent with predictions of the toy model: magnon–magnon damping is stronger and exhibits a quadratic temperature dependence, whereas magnon–phonon damping varies linearly with temperature. To assess the Gilbert or non-Gilbert character of the damping, we evaluate the dependence of the calculated FMR linewidth 
$\delta\varepsilon$ on the energy 
$\Delta$ of the ${\bf k}=0$ acoustic magnon mode. The results, shown in Fig.~\ref{fig:DFT}(c), are normalized to the linewidth at an external field of $0.25\,{\rm T}$. The linewidth arising from magnon–phonon scattering follows Eq.~(\ref{Gilb}) at low magnetic fields, indicating Gilbert-like damping. At higher magnetic fields it starts to deviate from Eq.~(\ref{Gilb}) as the phonon modes satisfying energy and momentum conservation become magnetic-field dependent. This dependence is especially strong for magnon-non-conserving (MNC) processes~\cite{SI}. By contrast, damping due to magnon–magnon scattering is distinctly non-Gilbert at all the magnetic fields and follows Eq.~(\ref{mag3D}) derived from the toy model.

Monolayer CrSBr is a ferromagnetic semiconductor that adopts an orthorhombic crystal structure belonging to the Pmmm space group. Within each layer, the Cr atoms reside in a distorted octahedral coordination environment. Along the $a$ axis, neighboring Cr atoms are linked through both sulfur and bromine atoms,  whereas along the b and c directions the Cr atoms are interconnected solely via sulfur atoms~\cite{CrSBr0,CrSBrwaves,CrSBrStrain}, as shown in the inset in Fig.~\ref{fig:DFT}(b). This figure shows that the temperature dependence of both damping mechanisms follows the predictions of the model: the magnon--magnon contribution to the damping scales as $\alpha \propto T^2$, whereas the magnon--phonon contribution is linear in $T$. The latter is shown for both a free-standing ${\rm CrSBr}$ monolayer (2D phonons) and a layer on a substrate (3D phonons, see Appendix~\ref{end:CrSBrphon} for the details). Over a broad range of experimentally relevant temperatures and magnetic fields, the magnon--magnon contribution dominates.
  Figure~\ref{fig:DFT}(d) presents $\delta\varepsilon(\Delta)$ dependence in a CrSBr monolayer. The damping originating from magnon–phonon scattering is Gilbert at low fields for both phonon dispersions considered, whereas the damping due to magnon–magnon scattering is non-Gilbert and follows Eq.~(\ref{mag2D}).

\begin{figure}
 	\centering
 	\includegraphics[width=3.0in]{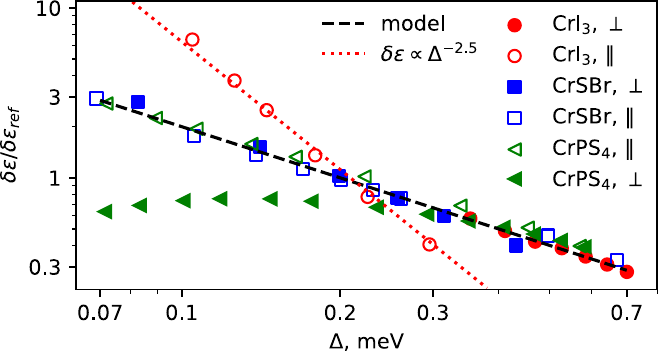}
 	\caption{ \label{fig:Many}
 		Damping due to MC 4-magnon processes in different materials compared to the toy model.    } 
\end{figure}

To assess the generality of Eq.~(\ref{mag2D}), which predicts strong non-Gilbert damping arising from magnon--magnon interactions in two-dimensional materials, we perform additional calculations for two other vdW magnetic insulators, ${\rm CrI_3}$~\cite{CrI3model,CrI3Tc,CrI32026} and ${\rm CrPS_4}$~\cite{longCrPS4,CrPS4Tc,CrPS4model}. In each case, we use spin Hamiltonians reported in the literature~\cite{CrI3model,CrPS4model}. For all three materials, including ${\rm CrSBr}$, we consider both parallel and perpendicular magnetization, as an external magnetic field can stabilize the magnetization along the hard axis, yielding six distinct configurations. 

Figure~\ref{fig:Many} compares the calculated damping with the predictions of the toy model. To enable this comparison, we rewrite Eq.~(\ref{mag2D}) in terms of the Curie temperature $T_c$ by using the molecular-field-theory relation between $T_c$,  $J$, and $S$~\cite{Aharoni}, leaving $\Delta_0$ as the only fitting parameter.~\cite{SI} We use the experimental Curie temperatures of ${\rm CrSBr}$, ${\rm CrI_3}$, and ${\rm CrPS_4}$, namely $132,{\rm K}$, $45,{\rm K}$, and $36,{\rm K}$, respectively~\cite{CrSBr0, CrI3Tc, CrPS4Tc}. Our results show that our model successfully reproduces the calculated damping in four of the six cases. In each of these cases, the fitted parameter $\Delta_0$ is comparable to the calculated energy of magnetic anisotropy~\cite{SI}, demonstrating that the damping can be estimated, at least to order of magnitude, from experimentally accessible quantities. The remaining two cases, ${\rm CrPS_4}$ with perpendicular magnetization and ${\rm CrI_3}$ with parallel magnetization, exhibit qualitatively different behavior. As discussed in the SI, both deviations arise from the strong wavevector dependence of the dipole--dipole interaction matrix elements, an effect that is neglected in the toy model. In particular, for ${\rm CrPS_4}$, destructive interference between the dipole--dipole interaction and the magnetocrystalline anisotropy (MCA) strongly suppresses the damping.

\section{Discussion and conclusions}

Our results demonstrate that strong non-Gilbert damping arising from magnon--magnon scattering is a generic feature of two-dimensional magnetic systems and is not suppressed by weak SOC alone. However, it can be strongly reduced by the combined effect of weak SOC and an external magnetic field: when the Zeeman energy exceeds $\Delta_0$, the damping coefficient scales as $\alpha \propto \Delta_0^2/B^2$. This behavior is observed for several \textit{ab initio}-derived spin Hamiltonians, indicating that the effect is present in many materials. Exceptions do exist, however, for example, due to destructive interference between MCA and dipole--dipole interactions.

At the level of the Boltzmann equation, this effect can be traced to small scattering matrix elements $W_{if}$
 being compensated by the large occupation numbers of low-energy thermal magnons. However, as shown in Appendix~\ref{end:afm}, the same mechanism persists in an antiferromagnetic toy model with a magnon dispersion and statistics, qualitatively different from the ones in ferromagnetic materials. We therefore suppose that the origin of this strong magnon-magnon damping might be more fundamental. In three-dimensional Heisenberg magnets, the ${\bf k} =0$ acoustic magnons are Nambu–Goldstone modes~\cite{Goldstone} associated with the spontaneous breaking of rotation symmetry of the magnetization or Néel vector. The stability of these modes implies that their damping arises only from weak anisotropic terms in the spin Hamiltonian induced by spin-orbit or dipole-dipole interaction. In contrast, long-range magnetic order is forbidden in two-dimensional Heisenberg magnets at finite temperature~\cite{Mermin}, rendering this argument inapplicable. In two dimensions, both the existence of magnetic order and the scattering of ${\bf k}=0$ acoustic magnons are controlled by anisotropy, leading to magnon–magnon damping that is effectively independent of the spin–orbit coupling strength.  Notably, in the ferromagnetic easy-plane version of the model, where the ${\bf k}=0$ mode remains gapless despite SOC, the scaling of $\delta\varepsilon(\Delta)$ is distinct and appears to be Gilbert-like at low $\Delta$.~\cite{SI}

Currently, even in high-quality YIG samples, the measured damping is typically attributed to residual impurities or substrate-induced stray fields~\cite{YIG-59,YIG-mk,YIG-ideal}, and is therefore considered extrinsic. In vdW materials, identifying intrinsic damping is even more challenging due to weak FMR signals~\cite{FMRvdW}, although existing experiments show similar trends~\cite{CrGeTe,Crplus}. Nevertheless, enhanced magnon damping arising from magnon--magnon interactions may limit the miniaturization of magnonic devices toward the monolayer regime. According to our results, this contribution can be distinguished in FMR experiments where the dependences of FMR linewidth on both magnon frequency and temperature are measured simultaneously. The magnon-magnon damping can be identified as a contribution proportional to $T^2$ and decreasing with $\Delta$ following Eq.~(\ref{mag2D}) or (\ref{mag3D}) for 2D and 3D materials, respectively. A possible mitigation strategy for this mechanism is the application of external magnetic fields, which significantly suppress the effective damping parameter $\alpha$ even at moderate fields. Alternatively, damping can be reduced by engineering the interplay between dipole--dipole interactions and MCA, which can be systematically explored using the \textit{MagnoFallas} code.

\section{Acknowledgements}

A.S. thanks Andrey Rybakov and Philipp Pirro for helpful discussions.

The authors acknowledge financial support from the European
Union (ERC-2021-StG-101042680 2D-SMARTiES), the
Spanish Government MCIU (PID2024-162182NAI00 2D-
MAGIC), the María de Maeztu Centre of Excellence Program
CEX2024-001467-M funded by MICIU/AEI/10.13039/
501100011033, and the Generalitat Valenciana (grant
CIDEXG/2023/1). A.M.R. thanks the Spanish
MIU (Grant No FPU21/04195).

\bibliography{damping}


\clearpage

\section{End Matter}

\subsection{General properties of the Boltzmann formalism}
\label{bol-Gen}

The most general expression for $\gamma_0$ within the Boltzmann formalism 
\begin{equation}\label{BolGen}
\gamma_0 = \frac{2\pi}{\hbar} \sum_{i,f}|W_{if}|^2    ({\cal N}_> - {\cal N}_<) \delta(E_i - E_f)
\end{equation}
involves a set of the initial states $i$ and the final states $f$ with one less ${\bf k} = 0$ acoustic magnon. $W_{if}$ is the matrix element of the Hamiltonian related to the transition from $i$ to $f$, ${\cal N}_>$ and ${\cal N}_<$ describe the occupation numbers of the quasi-particles (thermal phonons and magnons) relevant for the direct and reversed process, respectively. Dirac delta function 
$\delta(E_i - E_f)$ shows that the total energy of the initial state $E_i$ equals the energy of the final state $E_f$. 

Because of the energy conservation law, when $\Delta \ll T$ the term ${\cal N}_> - {\cal N}_<$ can be given by the expression~\cite{SI}:
\begin{equation}\label{BolGil}
{\cal N}_> - {\cal N}_< = \frac{\Delta}{T} \frac{\prod_A e^{\varepsilon_A/T} }{\prod_A \left(e^{\varepsilon_A/T}-1\right)\prod_C \left(e^{\varepsilon_C/T}-1\right)}
\end{equation}
Here index $A$ runs over all the quasi-particle anihilated in the process apart from the ${\bf k}=0$ acoustic magnon and $C$ denotes the quasi-particles created in the process.  $\varepsilon_A$ and $\varepsilon_C$ are the corresponding quasi-particle energies. Eq.~(\ref{BolGil}) shows that Eq.~(\ref{BolGen}) leads to a Gilbert damping when (I) the energies of all the participating quasi-particles are large compared to $\Delta$ and are not strongly modified by the magnetic field and (II) the magnetic field dependence of $W_{if}$ can be neglected. It happens in the case of magnon-phonon interaction (at least in weak magnetic fields) but not of magnon-magnon interaction which is controlled by low-energy magnons.

\begin{figure}[t]
 	\centering
 	\includegraphics[width=3.4in]{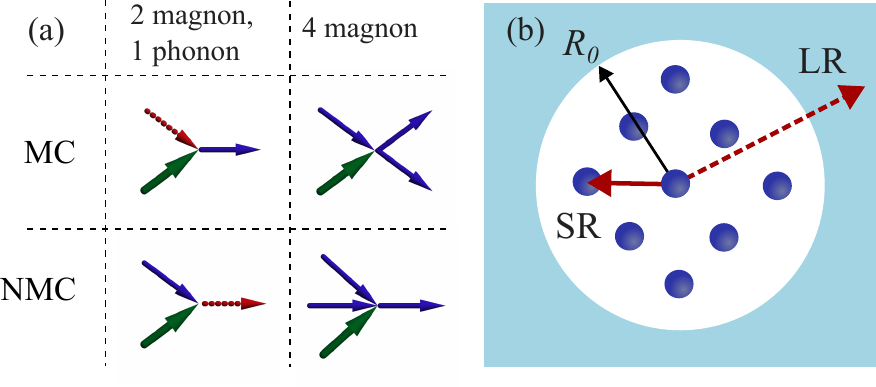}
 	\caption{ \label{fig:end1}
 		 		(a) The processes responsible for damping. Color code: green arrow - ${\bf k}=0$ acoustic magnon; blue arrows—thermal magnons; red arrows—phonons. (b) Illustration of the separation of the dipole–dipole interaction into short-range (SR) and long-range (LR) contributions.  } 
\end{figure}

In this work we consider four distinct processes leading to the relaxation of the ${\bf k}=0$ acoustic mode: MC and MNC 2-magnon 1-phonon and 4-magnon processes shown in Fig.~\ref{fig:end1}(a). All the four processes are included into the simulation of YIG and a monolayer of CrSBr, however, only MC processes are possible in the toy model. The detailed expression for $W_{if}$ relevant for all the processes are provided in SI~\cite{SI}. While 3-magnon processes are possible in the studied materials, the corresponding energy conservation laws require absorption of high-energy magnons leading to the contribution of 3-magnon processes being negligible at relevant temperatures. 

\begin{figure}
 	\centering
 	\includegraphics[width=3.4in]{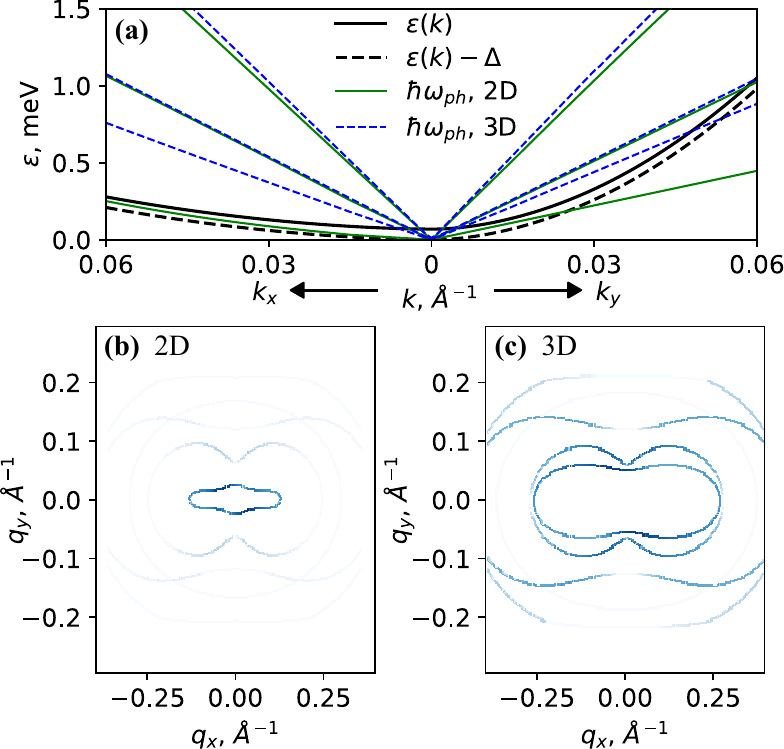}
 	\caption{ \label{fig:phoncon}
    (a) Low-energy magnon dispersion (black solid line) compared with the phonon dispersions obtained from \textit{ab initio} calculations for a CrSBr monolayer (green solid line) and bulk CrSBr (blue dashed line). The black dashed line indicates the phonon energy that must be absorbed in the MC process. (b) Contributions of phonons with different wavevectors to the MC magnon–phonon damping in the case of two-dimensional phonons. (c) Corresponding contributions for three-dimensional phonons. } 
\end{figure}

\subsection{Details of the numerical computations}
\label{numDet}

Although magnons in insulators can be viewed as collective electronic excitations~\cite{Green0, AliPh,Ali2}, we base our modeling on spin Hamiltonians derived from \textit{ab initio} calculations.

\textit{Ab initio} simulations of YIG and monolayer CrSBr were performed using VASP~\cite{VASP} and SIESTA~\cite{SIESTA} codes. Tight-binding Hamiltonians were constructed with wannier90~\cite{wannier90}, and spin Hamiltonians were derived via perturbation theory using TB2J~\cite{TB2J} software. The resulting magnon dispersions, calculated with RAD-tools package~\cite{RadTools}, agree well with previously reported results~\cite{YIGJex,CrSBrStrain}. Phonon calculations were carried out using the finite-displacement method implemented in phonopy~\cite{phonopy}. Spin–orbit coupling was included for CrSBr, whereas YIG was treated in the non-relativistic limit. Consequently, damping in YIG arises solely from dipole–dipole interactions added at the post-processing stage.

The damping coefficient $\alpha$ is computed numerically using our MagnoFallas package~\cite{MagnoFallas}. The package builds upon the RAD-tools code~\cite{RadTools}, which provides spin-Hamiltonian diagonalization and the interface with TB2J, and extends its functionality to evaluate magnon damping within the Boltzmann formalism. Below, we briefly outline the key elements of the underlying algorithms; a detailed description is provided in SI~\cite{SI}.

(I) Since $W_{if}$ in Eq.~(\ref{BolGil}) vanishes in the non-relativistic limit when energy and momentum conservation laws are fulfilled, but remains finite for arbitrary quasiparticle wave vectors, we evaluate $W_{if}$ only at points satisfying the conservation laws (\textit{scattering events}). To this end, the Brillouin zone is discretized into a $k$-grid. For each grid cell containing valid events, we select a representative point, compute $W_{if}$ for the corresponding wave vectors, and estimate the integral over the Dirac delta function using the marching-cubes method~\cite{marching}.

(II) Although dipole–dipole interactions can, in principle, be incorporated into Eq.~(\ref{SHgen}) as additional terms, this approach is practical only for a limited number of neighboring spins. Nevertheless, long-range dipole–dipole interactions play an important role in magnon dynamics and must be included in the calculations. To this end, we introduce a cutoff distance $R_0$ that separates the dipole–dipole interaction into short- and long-range contributions according to the interspin distance $r$
(Fig.~\ref{fig:end1}(b)). Short-range interactions ($r<R_0$) are treated explicitly within the spin Hamiltonian~(\ref{SHgen}). The long-range contribution is accounted for by coupling each spin to an averaged magnetization density, which allows for an analytical evaluation of its effect. Our treatment assumes spherically symmetric boundary conditions at large distances, corresponding to magnons near the center of a macroscopic magnetic sphere.

\begin{figure}
 	\centering
 	\includegraphics[width=3.4in]{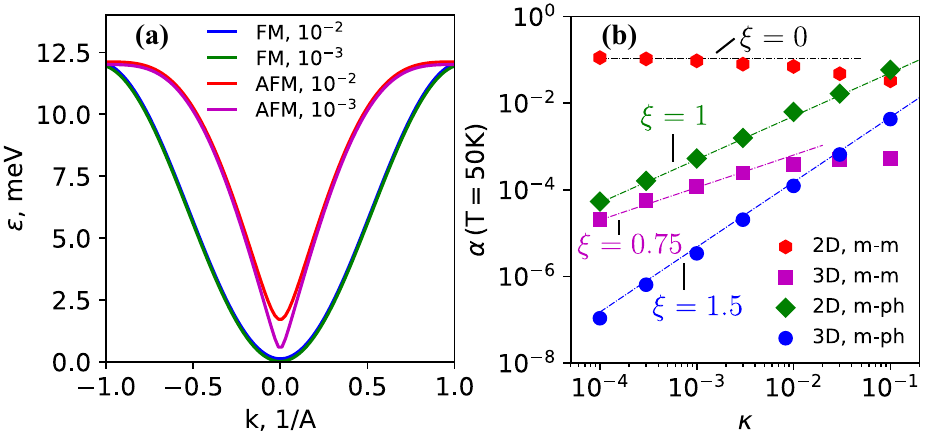}
 	\caption{ \label{fig:antimodel}
        (a) Comparison of the magnon dispersion of the ferromagnetic (FM) and antiferromagnetic (AFM) toy models with $\kappa = 10^{-2}$ and $10^{-3}$ as shown in legend. 
 		(b) Damping coefficient $\alpha$ at $T=50\,{\rm K}$ calculated for AFM toy model for different anisotropy parameters $\kappa$. The results are compared with power-law dependencies$\alpha \propto \kappa^\xi$.   } 
\end{figure}

\subsection{Difference between 2D and 3D phonons for the magnon-phonon damping mechanism in a monolayer of CrSBr}
\label{end:CrSBrphon}

The layers of vdW materials can be obtained either as free-hanging films of be put on a substrate. To simulate the difference between the two cases we calculate two versions of magnon-phonon damping in ${\rm CrSBr}$. In the first version (2D phonons) we calculate the phonons from the monolayer of ${\rm CrSBr}$. In the second version (3D phonons) we derive phonons from bulk ${\rm CrSBr}$, nevertheless we still consider a 2D spin Hamiltonian. The magnon-phonon damping in the two cases appears to be $\sim 10$ times different as shown in Fig.~\ref{fig:DFT}(b).

The pronounced difference between the two cases can be understood from the results shown in Fig.~\ref{fig:phoncon}. Figure~\ref{fig:phoncon}(a) compares the magnon dispersion with the dispersions of 2D and 3D phonons. In the 2D case, the phonon branch corresponding to the acoustic mode with atomic displacements perpendicular to the monolayer exhibits significantly lower energies, consistent with the presence of flexural phonons~\cite{Landau7}. As a result, crossings between the phonon and magnon spectra occur at much lower energies. Figures~\ref{fig:phoncon}(b) and (c) show the contributions of phonons with different wavevectors to the damping for 2D and 3D phonons, respectively. In the 2D case, the damping is dominated by low-wavevector flexural phonons, which have low energies and correspondingly large occupation numbers. In contrast, for 3D phonons the damping is governed by phonons with higher energies and smaller occupation numbers, resulting in an approximately 10-fold reduction of the damping parameter $\alpha$.

\subsection{antiferromagnetic toy model }
\label{end:afm}

Figure~\ref{fig:antimodel}(a) shows the magnon dispersion obtained from the toy model with antiferromagnetic exchange coupling $J=-2\,{\rm meV}$, $\kappa = 10^{-2}$ and $10^{-3}$, while all other parameters are the same as those used in Fig.~\ref{fig:model}. In contrast to the quadratic dispersion of the ferromagnetic toy model, the antiferromagnetic magnon spectrum becomes linear in $k$ at small $\kappa$.
Figure~\ref{fig:antimodel}(b) presents the numerically calculated damping parameter 
$\alpha$ for this model at $T=50\, {\rm K}$ and for different values of 
$\kappa$. As in the ferromagnetic case, magnon–magnon scattering in the two-dimensional model remains unsuppressed at low $\kappa$, resulting in a finite damping even in this limit.

\begin{widetext}

\newpage

\begin{center}
\normalfont\LARGE\bfseries SUPPORTING INFORMATION
\end{center}

\tableofcontents
    
\section{Derivation and General Properties of the Boltzmann Equation}

\subsection{Second-Quantized Description of Spins and Atom Displacements}

The spin Hamiltonian given in Eq.~(1) of the main text can describe both ferromagnetic, ferrimagnetic, and antiferromagnetic systems. In cases where the ground state is not strictly ferromagnetic, we first perform rotations of the local coordinate frames associated with each spin, following the procedure detailed in Ref.~\cite{spinW}. This transformation maps the system onto an effective ferromagnetic spin Hamiltonian, which can again be expressed in the general form of Eq.~(1) in the main text.

We then derive the corresponding magnon Hamiltonian using the Holstein–Primakoff transformation.
\begin{equation}\label{HP1}
S_+ \approx \sqrt{2S} a - \frac{a^+ a a}{\sqrt{2S}}, \quad S_- \approx \sqrt{2S} a^+ - \frac{a^+ a^+ a}{\sqrt{2S}}, \quad S_z = S-a^+a
\end{equation}
Here, $a$ and $a^+$ denote the magnon annihilation and creation operators associated with the spin $S$, respectively. The Cartesian spin components are expressed as $S_x = (S_+ + S_-)/2$ and $S_y = (S_+ - S_-)/2i$. In the Holstein–Primakoff expansion of $S_+$ and $S_-$, terms of fifth and higher order in magnon operators are neglected. For the calculation of magnon eigenstates and magnon–phonon interactions, only terms up to the second order are retained, whereas the treatment of magnon–magnon interactions includes contributions up to the fourth order.

The atomic displacements $u_k$
are expressed in terms of phonon operators as follows \cite{phonopy}.
\begin{equation}\label{phon}
u_k^{(\alpha)}({\bf R}) = \sum_{{\bf q},\lambda} C_{{\bf q}\lambda}^{(k,\alpha)}
\sqrt{\frac{\hbar}{2Nm_k \omega_{{\bf q}\lambda}}}
\left(
b_{{\bf q},\lambda}e^{i{\bf q}{\bf R}} + b_{{\bf q},\lambda}^+ e^{-i{\bf q}{\bf R}}
\right)
\end{equation}
Here, $u_k^{(\alpha)}({\bf R}) $ denotes the displacement of atom $k$ in the unit cell located at 
${\bf R}$ along the Cartesian direction $\alpha$.
The summation runs over phonon wavevectors  ${\bf q}$
and branches $\lambda$. ${\bf q}{\bf R}$ corresponds to the scalar product of vectors ${\bf q}$ and ${\bf R}$, here and below. The coefficients $C_{{\bf q}\lambda}^{(k,\alpha)}$ describe the displacement eigenvectors of atom $k$ for the corresponding phonon mode. $\omega_{{\bf q}, \lambda}$ is the phonon frequency, $m_k$ is the mass of atom $k$, and $N$ is the total number of unit cells in the sample. Finally, $b_{{\bf q},\lambda}$ and $b_{{\bf q},\lambda}^+$ are the phonon annihilation and creation operators, respectively.

\subsection{General Formulation of the Boltzmann Formalism}

The most general Boltzmann expression for the relaxation of 
${\bf k}=0$ acoustic magnons is given by
\begin{equation}\label{bolt0}
\frac{d N_0}{dt} = -\frac{2\pi}{\hbar} \sum_{if} |W_{if}|^2 \left[ N_0{\cal N}_{>} - (1+N_0) {\cal N}_{<}  \right] \delta(E_i - E_f).
\end{equation}
It involves a set of initial and final many-body states of the system, denoted by $i$ and $f$, respectively. The initial state contains $N_0$ acoustic magnons with ${\bf k} = 0$, while the final state contains one fewer ${\bf k}=0$ acoustic magnon. The two states may also differ in the occupation numbers of other quasiparticles, such as magnons or phonons; these differences are incorporated in Eq.~(\ref{bolt0}) through ${\cal N}_{>}$ and ${\cal N}_{<}$.
 $W_{if}$ denotes the matrix element of the Hamiltonian responsible for the transition, and the delta function $\delta(E_i - E_f)$ enforces energy conservation, requiring the total quasiparticle energy 
$E_i$ of the initial state to equal that of the final state $E_f$. We consider a relaxation of a classical spin wave with a vary large occupation number $N_0$. In this case the relaxation rate $\gamma_0$ can be expressed as
\begin{equation}\label{bolt0A}
\gamma_0 = \frac{2\pi}{\hbar} \sum_{if} |W_{if}|^2 \left[ {\cal N}_{>} - {\cal N}_{<}  \right] \delta(E_i - E_f).
\end{equation}
The part of Eq.~(\ref{bolt0A}) associated with the occupation numbers can be written explicitly as follows:
\begin{equation}\label{occ1}
{\cal N}_{>} -  {\cal N}_{<}  = \prod_{A} N_A \prod_{C}(1+N_C) - \prod_{A} (N_A + 1) \prod_{C}N_C
\end{equation}
The index 
$A$ labels the quasiparticles absorbed during the relaxation process, while the index 
$C$ labels those that are created. The occupation numbers $N_A$ and $N_C$
are both described by the Bose–Einstein distribution.
\begin{equation}\label{Boze}
N_{A,C} = \frac{1}{e^{\varepsilon_{A,C}/T} - 1},
\end{equation}
where, $\varepsilon_A$ and $\varepsilon_C$ are the energies of the absorbed and created quasi-particles, respectively. 

Eq.~(\ref{occ1}) can be reduced to 
\begin{equation}\label{occ2}
{\cal N}_> - {\cal N}_<  =  \frac{\exp\left( \frac{\sum_C \varepsilon_C}{T} \right) - 
\exp\left( \frac{\sum_A \varepsilon_A}{T}\right)}{\prod_A (e^{\varepsilon_{A}/T} - 1) \prod_C (e^{\varepsilon_{C}/T} - 1)}
\end{equation}
The energy conservation law results in the relation $\sum_C \varepsilon_C = \Delta + \sum_A \varepsilon_A$. When the energy of the ${\bf k}=0$ acoustic magnon is small compared to temperature, $\Delta \ll T$, the exponent $e^{\Delta/T}$ can be expanded into Taylor series: 
\begin{equation}\label{BolGil}
{\cal N}_> - {\cal N}_< \approx \frac{\Delta}{T} \frac{\prod_A e^{\varepsilon_A/T} }{\prod_A \left(e^{\varepsilon_A/T}-1\right)\prod_C \left(e^{\varepsilon_C/T}-1\right)}
\end{equation}
Eq.~(\ref{BolGil}) shows that the Boltzmann equation leads to a Gilbert damping when (I) the energies of all the participating quasi-particles are large compared to $\Delta$ and are not strongly modified by the magnetic field and (II) the magnetic field dependence of $W_{if}$ can be neglected. It happens in the case of magnon-phonon interaction (at least in weak magnetic fields) but not of magnon-magnon interaction which is controlled by low-energy magnons. 

In this work we consider four distinct processes leading to the relaxation of the ${\bf k}=0$ acoustic mode: magnon-conserving (MC) and magnon-non-conserving (MNC) 2-magnon 1-phonon and 4-magnon processes. All the four processes are included into the \textit{ab initio}-based simulations, however, only MC processes are possible in the toy model.  While 3-magnon processes are possible in some of the studied materials, the corresponding energy conservation laws require absorption of high-energy magnons leading to the contribution of 3-magnon processes being negligible at relevant temperatures (see Fig.~\ref{fig:SI-mag}). 

\subsection{Boltzmann Equation for Magnon–Phonon Interactions}

Because we focus on the relaxation of the ${\bf k}=0$ acoustic magnon, single-magnon–single-phonon processes are forbidden by momentum and energy conservation. We further assume that the magnon gap $\Delta$ is smaller than other magnon energies, which restricts the allowed two-magnon–one-phonon scattering channels to two distinct  processes. These processes are illustrated in Fig.~\ref{fig:SI-phon}(a,b).

In the magnon-conserving (MC) process, a ${\bf k}=0$ acoustic magnon absorbs a phonon with wavevector ${\bf q}$, acquires the same wavevector, and may undergo a transition between magnon branches. The magnon–non-conserving (MNC) process involves the recombination of the ${\bf k}=0$ acoustic magnon
 with another magnon of wavevector ${\bf k}_1$, accompanied by the emission of a phonon with wavevector ${\bf k}_1$. Figures~\ref{fig:SI-phon}(c,d) illustrate the corresponding energy and momentum conservation laws for the MC and MNC processes, respectively. Within the toy model [Eq.~(3) of the main text], only MC processes are allowed; however, for real materials, both types of processes are taken into account in this study.

\begin{figure*}
 	\centering
 	\includegraphics[width=3.5in]{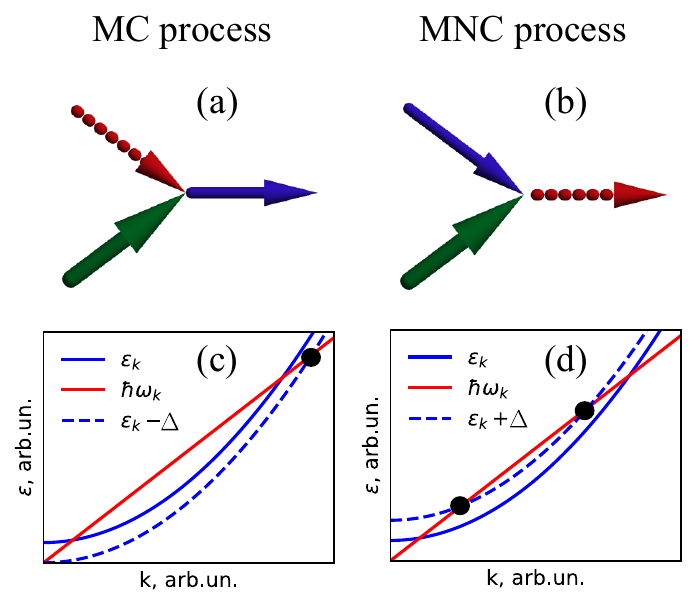}
 	\caption{ \label{fig:SI-phon}
    Two-magnon–one-phonon processes responsible for magnon damping. Panels (a,c) correspond to magnon-conserving (MC) processes, while panels (b,d) depict magnon–non-conserving (MNC) processes. Panels (a) and (b) show the participating quasiparticles: thick green arrows represent ${\bf k}=0$ acoustic magnons, blue arrows denote thermal magnons, and red dashed arrows indicate phonons. Panels (c) and (d) illustrate the corresponding energy–momentum conservation laws for the MC and MNC processes, respectively. The blue and red curves show the magnon and phonon dispersions, and the allowed scattering points in $(k,\varepsilon)$ space are indicated by black dots.
  } 
\end{figure*}

The Boltzmann expression for the ${\bf k} = 0$ acoustic magnon decay due to  the MC processes $\gamma_{m-ph}^{MC}$ is:
\begin{equation}\label{B-ph-MC}
\gamma_{m-ph}^{MC} = \frac{2\pi}{\hbar} \sum_{{\bf q, \lambda, \alpha}} \left|W_{\bf q, \lambda, \alpha}^{(m-ph, MC)}\right|^2\left(M_{{\bf q},\lambda} - N_{{\bf q},\alpha} \right) \delta(\Delta + \hbar \omega_{{\bf q}, \lambda} - \varepsilon_{{\bf q},\alpha})
\end{equation}
Here, the summation runs over phonon wavevectors ${\bf q}$, phonon branches $\lambda$  and bands of the resulting magnon $\alpha$. $W_{\bf q, \lambda, \alpha}^{(m-ph, MC)}$ is the matrix element of the Hamiltonian connecting the initial and final states ($W_{if}$ in Eq.~(\ref{bolt0})). It is derived by the second quantization of the initial spin and phonon Hamiltonian. $M_{{\bf q},\lambda}$ is the Bose–Einstein occupation number of the phonon and $\omega_{{\bf q},\lambda}$ is its frequency. $ N_{{\bf q},\alpha}$ is the thermal occupation number of the emitted magnon. $\varepsilon_{{\bf q},\alpha}$ is the energy of this magnon. 

According to the general expression (\ref{occ1}) at a reasonably high temperature $T \gg \Delta$, the term $M_{{\bf q},\lambda} - N_{{\bf q},\alpha}$ can be approximated as 
\begin{equation}\label{MNapp}
M_{{\bf q},\lambda} - N_{{\bf q},\alpha} \approx \frac{\Delta}{T} \frac{e^{\hbar \omega_{{\bf q}}/T}}{(e^{\hbar \omega_{{\bf q}}/T}-1)^2}
\end{equation}
leading to the Gilbert damping as long as the matrix element and energies $\hbar \omega_{{\bf q},\lambda}$ and $\varepsilon_{{\bf q},\alpha}$ are not significantly modified by the external magnetic field. 

A similar expression for magnon decay $\gamma_{m-ph}^{MNC}$ for the MNC processes is as follows
\begin{equation}\label{B-ph-MNC}
\gamma_{m-ph}^{MNC} = \frac{2\pi}{\hbar} \sum_{{\bf k}_1, \lambda, \alpha} \left|W_{{\bf k}_1, \lambda, \alpha}^{(m-ph, MNC)}\right|^2 \left(N_{{\bf k}_1,\alpha} - M_{{\bf k}_1,\lambda}\right) \delta(\Delta + \varepsilon_{{\bf k}_1,\alpha} - \hbar \omega_{{\bf k}_1, \lambda})
\end{equation}
Here ${\bf k}_1$ is the wavevector of magnon recombining with the ${\bf k}=0$ acoustic magnon and $W_{{\bf k}_1, \lambda, \alpha}^{(m-ph, MNC)}$ is the term in magnon-phonon Hamiltonian corresponding to the MNC process. 

\subsection{Boltzmann Equation for Magnon–Magnon Interactions}

Magnon–magnon interactions first appear at the level of three-magnon processes, illustrated in Fig.~\ref{fig:SI-mag}(a). In such a process, a 
${ \bf k}=0$ acoustic magnon absorbs another magnon with wavevector 
${\bf k}_1$, thereby increasing its energy. However, energy and momentum conservation permit this process only if the magnon undergoes a transition to a different band, which in turn requires magnons with very high energies (see Fig.~\ref{fig:SI-mag}(d)). We therefore neglect three-magnon processes in the present study. Instead, we include four-magnon magnon-conserving (MC) and magnon–non-conserving (MNC) processes, shown in Fig.~\ref{fig:SI-mag}(b,c). Within the toy model, only MC processes are allowed; however, both MC and MNC processes are considered for real materials.

\begin{figure*}
 	\centering
 	\includegraphics[width=5.0in]{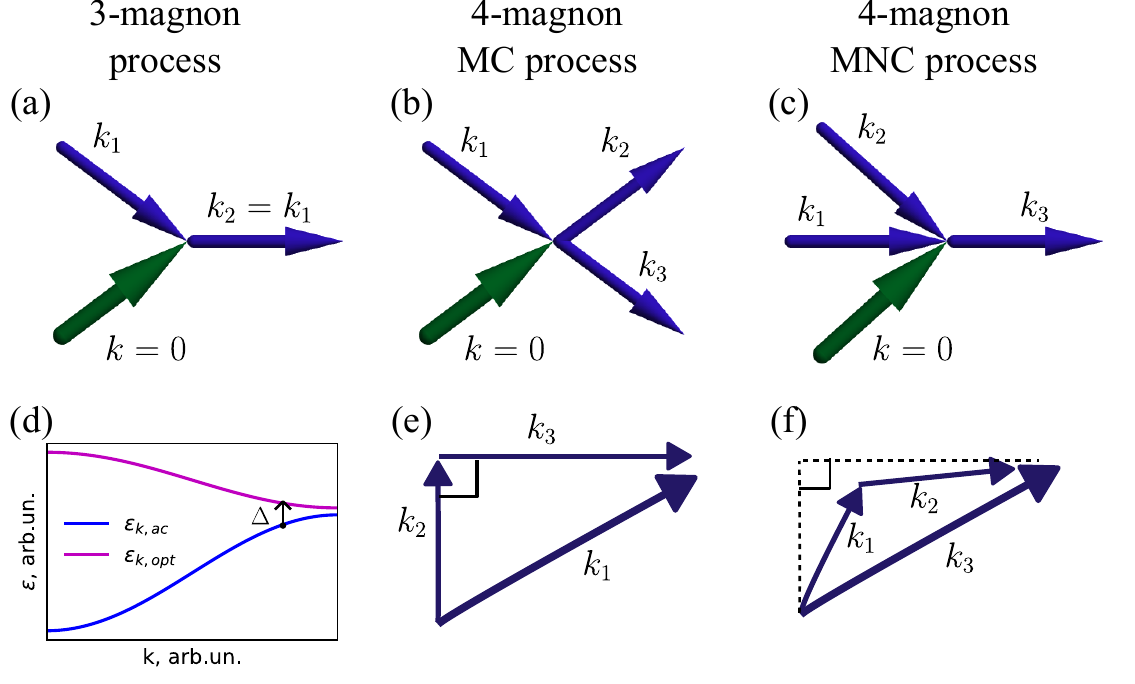}
 	\caption{ \label{fig:SI-mag}
 		Magnon-magnon interactions responsible for magnon damping. (a,d) 3-magnon processes, (b,e) 4-magnon MC processes, (c,f) 4-magnon MNC processes. Panels (a,b,c) show the interacting particles and their wavevectors. Panels (d,e,f) illustrate the energy-momentum conservation law.  } 
\end{figure*}

Due to the larger number of interacting particles, energy and momentum conservation are less restrictive for four-magnon processes. To illustrate this, we consider the MC process in which the initial 
${\bf k}=0$ acoustic magnon and another magnon with wavevector 
${\bf k_1}$ are annihilated, producing two magnons with wavevectors 
${\bf k}_2$ and ${\bf k}_3$. We focus on the case where all magnons remain in the acoustic band with relatively small wavevectors, allowing the use of a quadratic approximation for the magnon dispersion:
 \begin{equation} \label{smallE}
 \varepsilon({\bf k}) = \Delta + \varepsilon' k^2 , \quad \Delta =  \Delta_0 + \mu_b g B.
\end{equation}
Here $\Delta$ is the magnon gap dependent on the magnetic field $B$ which is assumed to be applied in z-direction;  $\varepsilon'$ is a constant (within the toy model it equals $\varepsilon' = |J|Sa^2$). The absorption of an acoustic magnon with an arbitrary wavevector ${\bf k}_1$ is possible. The wavevectors ${\bf k}_2$ and ${\bf k}_3$ should correspond to the momentum conservation law ${\bf k}_2 + {\bf k}_3 = {\bf k}_1$ and be perpendicular to each other (Fig.~\ref{fig:SI-mag}(e)) to fulfill the energy conservation law as long as Eq.~(\ref{smallE}) is relevant. 

Similar arguments applied to MNC processes lead to the momentum conservation law ${\bf k}_1 + {\bf k}_2 = {\bf k}_3$ and the energy conservation law $k_1^2 + k_2^2 = k_3^2 - 2 \Delta/\varepsilon'$. This means that the angle between vectors ${\bf k}_1$ and ${\bf k}_2$ is usually slightly larger than $90^{\circ}$. 

The expression for the magnon decay rate $\gamma_{4m}^{MC}$ related to the 4-magnon MC processes reads
\begin{multline}\label{B-4m-MC}
\gamma_{4m}^{MC} = \frac{2\pi}{\hbar} \sum_{{\bf k}_1, {\bf k}_2 \alpha_1, \alpha_2, \alpha_3} \left|W_{0,ac; {\bf k}_1, \alpha_1}^{{\bf k}_2, \alpha_2;{\bf k}_3, \alpha_3}\right|^2 \\
\times \left(N_{{\bf k_1},\alpha_1} + N_{{\bf k_1},\alpha_1}N_{{\bf k_2},\alpha_2} 
+N_{{\bf k_1},\alpha_1}N_{{\bf k_3},\alpha_3} 
- N_{{\bf k_2},\alpha_2}N_{{\bf k_3},\alpha_3}\right) \\
\times
 \delta(\Delta + \varepsilon_{{\bf k}_1,\alpha_1} - 
\varepsilon_{{\bf k}_2,\alpha_2} - \varepsilon_{{\bf k}_3,\alpha_3})
\end{multline}
Here, the summation runs over wavevectors ${\bf k}_1$ and ${\bf k}_2$ and the magnon band numbers $\alpha_1$, $\alpha_2$ and $\alpha_3$. It is assumed that ${\bf k}_3 = {\bf k}_1 - {\bf k}_2$.
$W_{0,ac; {\bf k}_1, \alpha_1}^{{\bf k}_2, \alpha_2;{\bf k}_3, \alpha_3}$ is the corresponding matrix element of the Hamiltonian. 

A similar expression for  $\gamma_{4m}^{MNC}$ related to the MNC 4-magnon processes is as follows
\begin{multline}\label{B-4m-MNC}
\gamma_{4m}^{MNC} = \frac{2\pi}{\hbar} \sum_{{\bf k}_1, {\bf k}_2 \alpha_1, \alpha_2, \alpha_3} \left|W_{0,ac; {\bf k}_1, \alpha_1; {\bf k}_2, \alpha_2}^{{\bf k}_3, \alpha_3}\right|^2 \\
\times \left(
 N_{{\bf k_1},\alpha_1}N_{{\bf k_2},\alpha_2} 
- N_{{\bf k_1},\alpha_1}N_{{\bf k_3},\alpha_3}
- N_{{\bf k_2},\alpha_2}N_{{\bf k_3},\alpha_3}
- N_{{\bf k_3},\alpha_3} \right) \\
\times
 \delta(\Delta + \varepsilon_{{\bf k}_1,\alpha_1} + 
\varepsilon_{{\bf k}_2,\alpha_2} - \varepsilon_{{\bf k}_3,\alpha_3})
\end{multline}
Here, $W_{0,ac; {\bf k}_1, \alpha_1; {\bf k}_2, \alpha_2}^{{\bf k}_3, \alpha_3}$ is the term in magnon Hamiltonian describing the anihilation of magnons  $(0,ac)$, $({\bf k}_1,\alpha_1)$, and $({\bf k}_2,\alpha_2)$ and the creation of the magnon $({\bf k}_3,\alpha_3)$. ${\bf k_3} = {\bf k}_1 + {\bf k}_2$ is assumed. 

The general expression (\ref{occ1}) is still valid for the 4-magnon processes. In particular in Eq.~(\ref{B-4m-MC}) the part related to the thermal occupation numbers can be reduced to
\begin{equation}\label{4Mapp}
\left(N_{{\bf k_1},\alpha_1} + N_{{\bf k_1},\alpha_1}N_{{\bf k_2},\alpha_2} 
+N_{{\bf k_1},\alpha_1}N_{{\bf k_3},\alpha_3} 
- N_{{\bf k_2},\alpha_2}N_{{\bf k_3},\alpha_3}\right) 
\approx \frac{\Delta}{T}
\frac{e^{\varepsilon_{{\bf k}_1, \alpha_1}/T}}{(e^{\varepsilon_{{\bf k}_1, \alpha_1}/T} - 1)
(e^{\varepsilon_{{\bf k}_2, \alpha_2}/T} - 1)
(e^{\varepsilon_{{\bf k}_3, \alpha_3}/T} - 1)}
\end{equation}
This equation would lead to the Gilbert-like damping if all the energies of all the participating magnons would be much larger than magnon gap.   Nevertheless, as it is shown in Fig.~1 of the main text, the most significant contribution to $\gamma_{4m}^{MC}$ comes from the magnons with very small  ${\bf k}_1$ wavevectors and with  $\varepsilon_{{\bf k}_1, \alpha_1} \sim \Delta$. It makes the magnon decay due to the 4-magnon processes non-Gilbert.

\section{Damping in the Minimal Toy Model}

\subsection{Magnon–Phonon Contribution to the Damping}

Within our toy model, the modifications of the spin Hamiltonian due to the atomic displacements, transformed to the magnon notations with expression (\ref{HP1}) are:
\begin{equation}\label{S-ph1}
H = \sum_{i;b} J' S ({\bf u}_{j(b)} - {\bf u}_i)\cdot {\bf e}_b
\left[\left(a_i^+ a_{j(b)} + a_i a_{j(b)}^+ - a_i^+a_i - a_{j(b)}^+a_{j(b)}  \right) - \kappa (a_i^+a_i + a_{j(b)}^+a_{j(b)}) \right]
\end{equation}
Here $i$ enumerates the spins and $b$ --- exchange interactions (bonds). ${\bf u}_i$ is the displacement of atom $i$, ${\bf e}_b$ is the direction of the bond $b$. This Hamiltonian conserves the number of magnons and prohibits MNC processes. To get the Hamiltonian matrix element 
$W_{\bf q, \lambda, \alpha}^{(m-ph, MC)}$  describing an MC process we use the second quantization expression (\ref{phon}):
\begin{equation}\label{modelWphon}
W_{\bf q, \lambda, \alpha}^{(m-ph, MC)} = -2i\kappa S J' \left( \frac{\hbar}{2Nm\omega_{{\bf q}{\lambda}}} \right)^{1/2}
 \sum_{b}  ({\bf e}_{\lambda} \cdot  {\bf e}_b)  \sin({\bf q} \cdot {\bf r}_b)
\end{equation}
Here ${\bf e}_{\lambda}$ is the displacement vector corresponding to the mode $\lambda$. 

To analytically calculate $\gamma_{m-ph}^{MC}$ we note that in Eq.~(\ref{B-ph-MC}) with the matrix element (\ref{modelWphon})
the dependence on the phonon direction and branch is present only in the term $({\bf e}_\lambda{\bf e}_b ) \sin({\bf k} {\bf r}_b) \approx ({\bf e}_\lambda{\bf e}_b ) ({\bf k}{\bf r_b})$. This allows us to make the summation over $\lambda$ and $b$ and, simultaneously, the integration over directions of ${\bf q}$ as follows:
\begin{equation}\label{int1}
\int do \sum_{\lambda} \left|\sum_b ({\bf e}_\lambda \cdot   {\bf e}_b ) ({\bf k} \cdot {\bf r_b}) \right|^2 = \sum_\lambda \int do \sin^2({\bf k}\cdot {\bf e}_\lambda a ) = D \cdot {\cal J}_{D}(ka),
\end{equation}
where $D$ is the system dimensionality. 

In Eq.~(\ref{int1}) we took into account that for each phonon branch $\lambda$ there is exactly one bond with ${\bf e}_b = {\bf e}_\lambda$ and all other bonds have the vectors ${\bf e}_b$ orthogonal to ${\bf e}_\lambda$.
We give an explicit expression for ${\cal J}_{3D}$ while ${\cal J}_{2D}$ is approximated in the limit of $ka \ll 1$.
\begin{equation}\label{J3}
{\cal J}_{3D}(ka) = \frac{2\pi(ka - \cos(ka)\sin(ka))}{ka} \approx \frac{4\pi (ka)^2}{3},
\end{equation}
\begin{equation}\label{J2}
{\cal J}_{2D}(ka) = \int_0^{2\pi} \sin^2(ka\cos(\phi))d\phi \approx \pi (ka)^2.
\end{equation}

 With Eq.~(\ref{int1}) it is possible to reduce Eq.~(\ref{B-ph-MC}) to the expressions (4,5) from the main text.

\subsection{Magnon–Magnon Contribution to the Damping}

Within the toy model the NMC 4-magnon processes are impossible while the matrix element of the 4-magnon Hamiltonian describing a MC 4-magnon process can be given with a simple expression 
\begin{equation}\label{4Mmel}
W_{0,ac; {\bf k}_1, ac}^{{\bf k}_2, ac;{\bf k}_3, ac} = \frac{2 \Delta_0}{NS}.
\end{equation}
as long as the conservation laws are fulfilled and spin-orbit interaction is small.  $W_{0,ac; {\bf k}_1, ac}^{{\bf k}_2, ac;{\bf k}_3, ac}$ described with Eq.~(\ref{4Mmel}) does not depend on the wavevectors of the participating magnons.

\begin{figure*}
 	\centering
 	\includegraphics[width=5.0in]{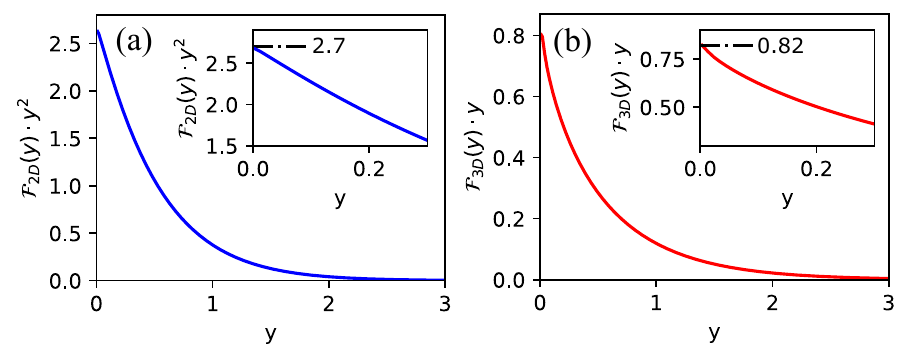}
 	\caption{ \label{fig:F2F3}
 		The function ${\cal F}_{2D}$ multiplied by $y^2$ (a) and the function ${\cal F}_{3D}$ multiplied by $y$ (b)
        calculated numerically. Insets show the functions at small values of $y$ and $y \rightarrow 0$ limits.  } 
\end{figure*}

Eq.~(\ref{4Mmel}) allows the analytical calculation of $\gamma_{4m}^{MC}$ within the toy model. Below we provide the expression for $\alpha_{4m}^{MC} = \hbar \gamma_{4m}^{MC}/\Delta$ in 2D case:
\begin{equation}\label{a4Mmodel0}
\alpha_{4m}^{MC; 2D} =  \frac{1}{2} \frac{a^4 \Delta_0^2}{(2\pi \epsilon' S)^2}
{\cal F}_{2D}(\Delta/T),
\end{equation}
where
\begin{equation}\label{F2d}
{\cal F}_{2D}(y) = \int_0^{\infty} dx \frac{e^{x +y}}{e^{x+y}-1} \int_{-\pi/2}^{\pi/2} \frac{d\theta}{(e^{y+x\cos^2\theta}-1)(e^{y+x\sin^2\theta}-1)}
\end{equation}
The function $F_{2D}(y)$, calculated numerically, is shown in Fig.~\ref{fig:F2F3}(a). For small values of $y$ (which corresponds to $T \gg \Delta$) it can be approximated as ${\cal F}_{2D}(y) \approx 2.7/y^2$ leading to the expression
\begin{equation}\label{4Malp2D-NG1}
\alpha_{4m}^{MC; 2D} =  \frac{1.35 T^2 a^4 \Delta_0^2}{(2 \pi S\varepsilon')^2\Delta^2}, \quad \gamma_{4m}^{MC; 2D} = \frac{1.35 a^4 \Delta_0^2 T^2}{\hbar(2\pi S \varepsilon')^2} \frac{1}{\Delta}
\end{equation}
The inverse proportionality $\gamma_{4m}^{MC; 2D} \propto 1/\Delta$ clearly shows non-Gilbert character of the damping. 

The similar calculations can also be made in 3D case leading to
\begin{equation}\label{a4M3Dmodel0}
\alpha_{4m}^{MC; 3D} =  \frac{1}{(2\pi)^3} \frac{\Delta_0^2 T a^6}{S^2(\varepsilon')^3} {\cal F}_{3D}(\Delta/T),
\end{equation}
with the function ${\cal F}_{3D}(y)$ equal to
\begin{equation}\label{F3D}
{\cal F}_{3D}(y) = \int_0^{\infty} xdx \frac{e^{y+x}}{e^{y+x} - 1} \int_0^1 dc \frac{c}{(e^{y+x c^2}-1)(e^{y+x(1-c^2)}-1)}
\end{equation}
This function is shown in Fig.~\ref{fig:F2F3}(b) and can be approximated for small values of $y$ as follows:
\begin{equation}\label{F3Dapr}
{\cal F}_{3D}(y) \approx \frac{0.82}{y} - 2.94
\end{equation}

 It leads to:
\begin{equation}\label{NG3r}
\alpha_{4m}^{MC; 3D} = \frac{\Delta_0^2 T^2 a^6}{(2\pi S \varepsilon')^3 \Delta}\left(0.82 - 2.94 \frac{\Delta}{T} \right), \quad
\gamma_{4m}^{MC; 3D} = \frac{\Delta_0^2 T^2 a^6}{\hbar(2\pi S \varepsilon')^3 }\left(0.82 - 2.94 \frac{\Delta}{T} \right)
\end{equation}
Similarly to the 2D case the relaxation is non-Gilbert.

\subsection{Antiferromagnetic toy model }
\label{end:afm}

\begin{figure}
 	\centering
 	\includegraphics[width=5.0in]{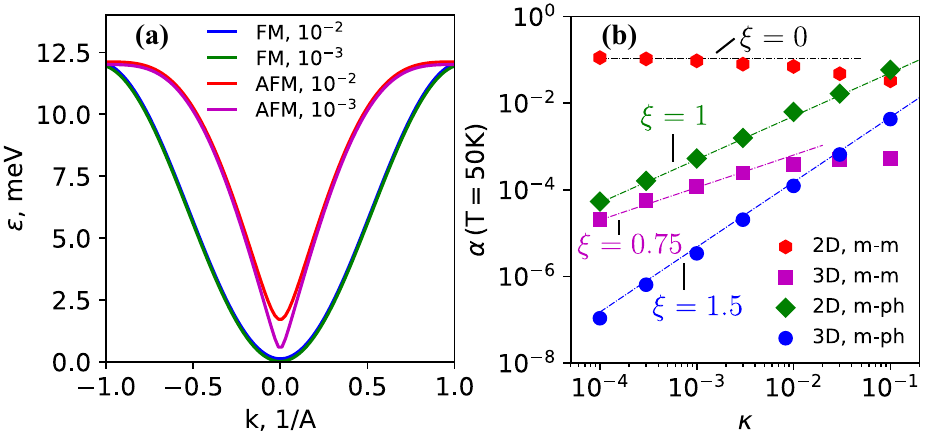}
 	\caption{ \label{fig:antimodel}
        (a) Comparison of the magnon dispersion of the ferromagnetic (FM) and antiferromagnetic (AFM) toy models with $\kappa = 10^{-2}$ and $10^{-3}$ as shown in legend. 
 		(b) Damping coefficient $\alpha$ at $T=50\,{\rm K}$ calculated for AFM toy model for different anisotropy parameters $\kappa$. The results are compared with power-law dependencies$\alpha \propto \kappa^\xi$.   } 
\end{figure}

Figure~\ref{fig:antimodel}(a) shows the magnon dispersion obtained from the toy model with antiferromagnetic exchange coupling $J=-2\,{\rm meV}$, $\kappa = 10^{-2}$ and $10^{-3}$, while all other parameters are the same as those used in the main text. In contrast to the quadratic dispersion of the ferromagnetic toy model, the antiferromagnetic magnon spectrum becomes linear in $k$ at small $\kappa$.
Figure~\ref{fig:antimodel}(b) presents the numerically calculated damping parameter 
$\alpha$ for this model at $T=50\, {\rm K}$ and for different values of 
$\kappa$. As in the ferromagnetic case, magnon–magnon scattering in the two-dimensional model remains unsuppressed at low $\kappa$, resulting in a finite damping even in this limit.

\subsection{Easy-plane toy model}

\begin{figure*}
 	\centering
 	\includegraphics[width=6.0in]{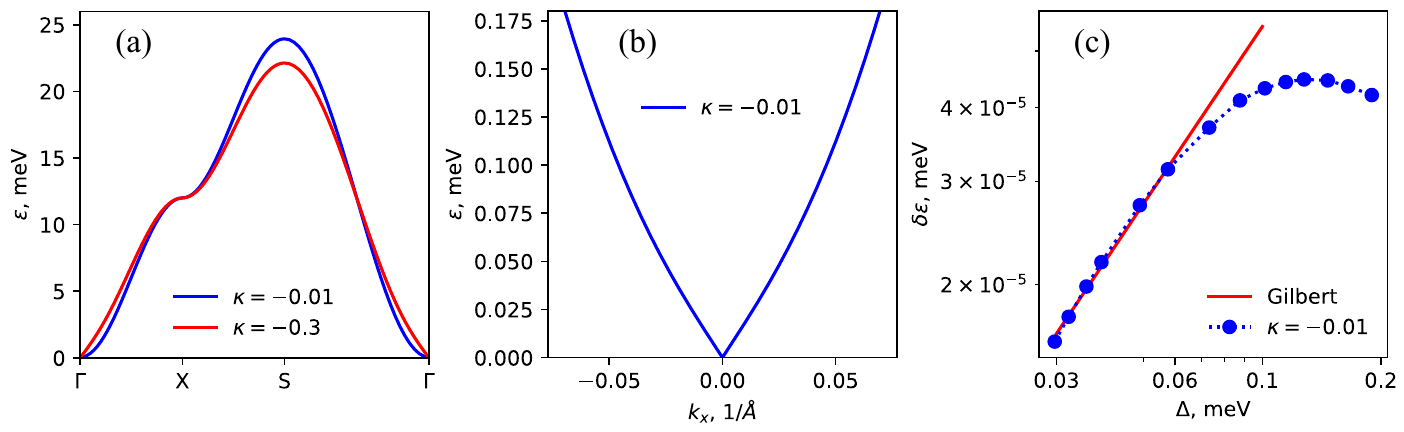}
 	\caption{ \label{fig:SI-XXZ}
 		Simulation of magnon dispersion and magnon-magnon damping in the easy-plane model. (a) Magnon dispersion in the toy model with $\kappa=-0.01$ and $\kappa = -0.3$. (b) Magnon dispersion in the toy model with $\kappa = -0.01$ for small values of ${\bf k}$. (c) Magnon linewidth due to MC magnon-magnon processes in the toy model with $\kappa = -0.01$ compared with the Gilbert damping.  } 
\end{figure*}

The results presented in the main text correspond to positive values of $\kappa$, which describe the easy-axis anisotropy regime. For $\kappa < 0$, the model instead exhibits easy-plane anisotropy. This case is  different, because the ${\bf k}=0$ acoustic magnons remain Goldstone modes even in the presence of spin--orbit coupling. As a consequence, the magnon dispersion is linear at low momenta in the absence of an external magnetic field~\cite{easyXXZ}. Figure~\ref{fig:SI-XXZ} summarizes the results of our numerical simulations of the magnon dispersion and the magnon--magnon damping arising from the MC processes in this regime. Figure~\ref{fig:SI-XXZ}(a) shows the magnon dispersion for the toy model with $\kappa=-0.01$ and $\kappa=-0.3$. For $\kappa=-0.3$, the dispersion is clearly linear in the vicinity of the $\Gamma$ point. In contrast, the linear behavior is less apparent on the same scale for $\kappa=-0.01$. Nevertheless, as demonstrated in Fig.~\ref{fig:SI-XXZ}(b), the dispersion also becomes linear in this case for sufficiently small wave vectors ${\bf k}$. Figure~\ref{fig:SI-XXZ}(c) shows the FMR linewidth $\delta\varepsilon$ as a function of $\Delta$ for $\kappa=-0.01$, numerically calculated for the MC magnon-magnon scattering. The results are compared with the Gilbert damping. Interestingly, at low magnetic fields, the magnon--magnon damping exhibits Gilbert-like behavior in the easy-plane toy model.

\section{Principles of the \textit{MagnoFallas} code}

The damping coefficients $\alpha$ are computed numerically using our MagnoFallas package~\cite{MagnoFallas}. The package builds upon the RAD-tools code~\cite{RadTools}, which provides spin-Hamiltonian diagonalization and the interface with TB2J, and extends its functionality to evaluate magnon damping within the Boltzmann formalism. 
The code is available at \url{https://github.com/AndreiShumilin/MagnoFallas}

Here we describe the general physical and mathematical framework implemented in this code.

\subsection{Magnon Energies}

We take the spin Hamiltonian obtained from the TB2J software to represent the exchange interaction Hamiltonian of the material. The corresponding magnon energies are computed using conventional linear spin-wave theory \cite{spinW}, as implemented in the code.

However, a complete description of the magnon excitations requires introduction of the contributions from the dipole–dipole interaction. In our code, the dipole–dipole interaction is separated into short-range (SRDD) and long-range (LRDD) components by introducing a cutoff distance $R_0$. The SRDD contribution is incorporated into the model as an additional term in the spin Hamiltonian [Eq.~(1) of the main text].

\begin{equation}\label{Hdd0}
H_{SRDD} = C_{dd}\sum_{i,j,v} g_i g_j  {\bf S}_i(0) \widehat{D}_{ij}(v) {\bf S}_j(v), \quad \widehat{D}_{ij}(v) \equiv D({\bf r}_{i,j,v})
\end{equation}
Here each term denoted by the indicies $i,j,{\bf v}$ describes the dipole-dipole interaction between magnetic atoms $i$ and $j$ in different unit cells related by an integer vector ${\bf v}$ (a vector ${\bf v} = (0,0,0)$ corresponds to the interaction of atoms $i$ and $j$ in the same unit cell). 
 $C_{dd} = \mu_0 \mu_B^2/4\pi$ is a constant describing the dipole-dipole interaction of the two magnetic moments equal to Bohr magneton $\mu_B$ at a unit distance.  The matrix $\widehat{D}_{ij}(v)$ depends only on the real space vector ${\bf r_{i,j,{\bf v}}}$ between spins $(i,0)$ and $(j,v)$, which is equal to  ${\bf r}_i^{(0)} - {\bf r}_j^{(0)} + v[1]{\bf a}_1 + v[2]{\bf a}_2 + v[3]{\bf a}_3$. Here  ${\bf a}_1$, ${\bf a}_2$ and ${\bf a}_3$ are the lattice vectors of the material.  Therefore,  $\widehat{D}_{ij}(v)  = \widehat{D}({\bf r_{i,j,{\bf v}}})$ with
\begin{equation}\label{Ddef}
\widehat{D}({\bf r}) = \frac{1}{|r|^3} \left(
\begin{array}{ccc}
  3e_x e_x - 1 & 3e_x e_y & 3e_x e_z \\
  3e_y e_x & 3e_y e_y - 1 & 3e_y e_z \\
  3e_z e_x & 3e_z e_y &  3e_z e_z - 1
\end{array}
\right),
\end{equation}
where $(e_x,e_y,e_z) ={\bf r}/|r|$ is the direction of ${\bf r}$. 

\begin{figure*}
 	\centering
 	\includegraphics[width=5.0in]{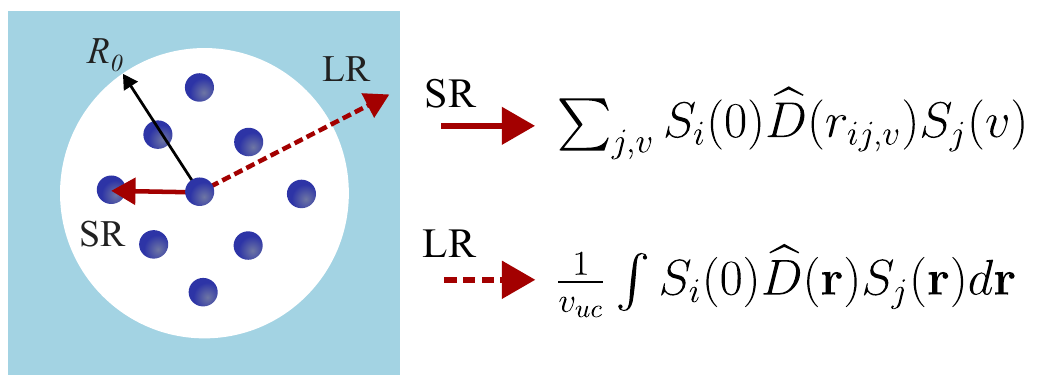}
 	\caption{ \label{fig:ddSI}
 		Illustration of the separation of the dipole–dipole interaction into short-range (SRDD) and long-range (LRDD) contributions. } 
\end{figure*}

To incorporate the LRDD contribution, we consider an “initial” spin of type $i$ located in the 0th unit cell and assume that spins of type $j$ at distances $r>R_0$ can be described by a continuous spin density ${\bf S}({\bf r})/v_{uc}$, where $v_{uc}$ is the unit-cell volume. The orientation of ${\bf S}({\bf r})$ may correspond either to a uniform magnetization or to a spin-wave excitation. The resulting contribution to the Hamiltonian is given by: 
\begin{equation}\label{HLRDD}
H_{LRDD} = C_{dd}\sum_{i,j} g_i g_j  \frac{1}{v_{uc}} \int_{|r|>R_{dd}}  {\bf S}_i(0) \widehat{D}({\bf r}) {\bf S}_j({\bf r}) d{\bf r}
\end{equation}

Both the magnon energies and the four-magnon interaction require evaluating integrals of the matrix $\widehat{D}({\bf r})$ over the region  $|r|>R_{dd}$, with the exponent ${\rm exp}(i{\bf k}{\bf r})$. Here, ${\bf k}$ denotes either a magnon wave vector or a combination (difference) of the wave vectors of interacting magnons. Accordingly, we introduce
\begin{equation}\label{Dkdef}
\widehat{D}({\bf k}; R_{0}) = \int_{|r|>R_{0}} \widehat{D}({\bf r}) e^{i{\bf k}{\bf r}} d{\bf r}
\end{equation}
This integral is taken semi-analitically. In particular, in 3D systems:
\begin{equation}\label{Dk3D}
D({\bf k}; R_{0}) = T^T D_0(|k|,R_{0}) T, \quad D_0(k,R_{0}) = 8\pi
 \left(
\begin{array}{ccc}
   - \frac{1}{2}\widetilde{D}_{3D}(|k|R_{0}) & 0 & 0 \\
  0 & - \frac{1}{2}\widetilde{D}_{3D}(|k|R_{0}) & 0 \\
  0 & 0 &  \widetilde{D}_{3D}(|k|R_{0})
\end{array}
\right),
\end{equation}
\begin{equation}\label{Dk3Da}
\widetilde{D}_{3D}(x) = \frac{x\cos(x) - \sin(x)}{x^3}, \quad x>0;
\end{equation}
Here, $T$ is a rotation matrix that aligns ${\bf k}$ with the z-direction. In Eqs.~(\ref{Dkdef}–\ref{Dk3Da}), we implicitly assume that the integration is performed over all angles and over radial distances $r$ ranging from $R_0$ to a large cutoff $R_{inf}$. We subsequently take the limit $R_{inf} \rightarrow \infty$. These boundary conditions correspond to a description of magnons located at the center of a macroscopically large spherical magnetic sample.

We also derive a similar expression for the 2D case:
\begin{equation}\label{Dk2D}
D({\bf k}; R_{0}) = {T}^T D_0^{(2D)}(|k|,R_{0}) {T}, \quad D_0(k,R_{0}) =  \frac{1}{R_0}
 \left(
\begin{array}{ccc}
   I_{x}^{(2D)}(|k|R_{0}) & 0 & 0 \\
  0 & I_{y}^{(2D)}(|k|R_{0}) & 0 \\
  0 & 0 &  I_{z}^{(2D)}(|k|R_{0})
\end{array}
\right),
\end{equation}
\begin{eqnarray}\label{Dk2Da-Ixyz}
I_x(x) =  (\pi/x)\left[2x^2  + J_1(x) (2 + 2x^2 - \pi x^3 H_0(x)) + xJ_0(x)(\pi x^2 H_1(x) - 2(1 + x^2)) \right]  \nonumber    \\
I_y(x) = 2\pi J_1(x)/x \\
I_z(x) = \pi x \left[ 2 - \frac{2(1+x^2)J_0(x)}{x} +  J_1(x) (2-\pi x H_0(x)) + \pi x J_0(x)H_1(x)\right] \nonumber
\end{eqnarray}

Here $J_n(x)$ is the Bessel function of order $n$ and $H_n(x)$ is the Struve function of the same order.

To incorporate the LRDD contribution into the magnon theory, we perform the Holstein–Primakoff transformation using the Fourier components $\mathbf{S}_i({\bf k}) = \int \mathbf{S}_i({\bf R}) e^{i{\bf k}{\bf R}}$. The resulting creation and annihilation operators are interpreted as magnon operators associated with magnetic atoms of type $i$
and wave vector ${\bf k}$. The Hamiltonian expressed in these terms is then treated within the conventional linear spin-wave theory workflow \cite{spinW}.

\begin{figure*}
 	\centering
 	\includegraphics[width=7.0in]{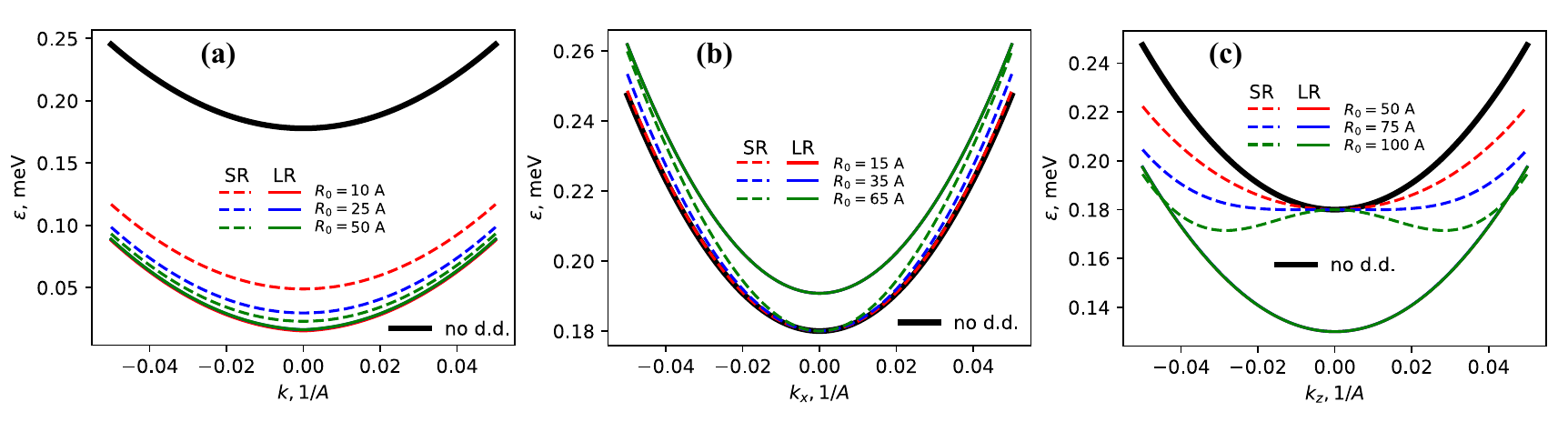}
 	\caption{ \label{fig:SI-ddEn} 
    Effect of the dipole–dipole interaction on magnon energies in the two-dimensional (a) and three-dimensional (b,c) toy models. Panel (b) correspond to the magnons propagating perpendicular to the magnetization direction while (c) shows magnons propagating in this direction. Model parameters are identical to those used in the main text. An external magnetic field of $B=0.5\,{\rm T}$ is applied in the 2D case. The black line denotes magnon energies calculated without the dipole–dipole interaction, while the dashed lines show results including only the short-range (SR) contribution. The colored solid lines correspond to calculations including both short-range and long-range (LR) contributions. Different colors indicate different cutoff distances $R_0$, as specified in the legend.
 	   } 
\end{figure*}

To validate our approach and assess the relative importance of SRDD and LRDD, we compare the magnon spectra of the toy model described in the main text with those obtained from the same model augmented by the dipole–dipole interaction. The results are shown in Fig.~\ref{fig:SI-ddEn}. We consider three cases: (i) a two-dimensional model with magnetization perpendicular to the film, stabilized by an external magnetic field of $0.5\,{\rm T}$ [Fig.~\ref{fig:SI-ddEn}(a)]; (ii) a three-dimensional model with magnetization along the $z$ direction and magnons propagating along the $x$ axis [Fig.~\ref{fig:SI-ddEn}(b)]; and (iii) the same three-dimensional model with magnons propagating along the $z$ axis [Fig.~\ref{fig:SI-ddEn}(c)].

In all panels, the thick black line represents the magnon dispersion of the original model without the dipole–dipole interaction. Colored lines correspond to numerical results obtained with different cutoff distances $R_0$. Dashed lines indicate calculations including only the SRDD contribution, while solid lines show results with both SRDD and LRDD contributions included.

Figure~\ref{fig:SI-ddEn}(a) demonstrates that magnon energies in the 2D model can be accurately estimated using only the SRDD contribution, provided that a sufficiently large cutoff $R_0$ is employed. However, when the LRDD contribution is included, comparable accuracy can be achieved with a much smaller cutoff $R_0$, thereby substantially reducing the required number of terms in the spin Hamiltonian. In contrast, as shown in Fig.~\ref{fig:SI-ddEn}(b,c), even relatively large cutoff values in the 3D case yield an accurate description of magnons only above a finite wave vector ${\bf k}$. This limitation arises from the very slow convergence of the dipole–dipole interaction integrals in 3D systems.

\subsection{Phonon Energies and Magnon–Phonon Coupling}

The phonon dynamical matrices are computed using the \textit{phonopy} software package \cite{phonopy}, and the results are imported via the provided API.

The MagnoFallas code implements two methods for calculating the magnon–phonon interaction, arising from exchange coupling and dipole–dipole coupling. The goal of both approaches is to evaluate the matrices 
$\partial \hat{J}_{ij}/\partial{\bf r}_k$ defined in the main text. The coordinate dependence of the exchange interaction  $\hat{J}_{ij}$ is obtained from a set of spin Hamiltonians calculated for arbitrary atomic displacements. It is assumed that the exchange interaction between spins $i$ and $j$ depends only on the distance between them, which is a reasonable approximation for magnon interactions with acoustic phonons. The derivative with respect to this distance of each element of the matrix  $\hat{J}_{ij}$
is estimated independently by performing a linear regression on the available data for exchange interactions and atomic displacements.

The displacement dependence of the dipole–dipole interaction is introduced through the derivative of the matrix
$\widehat{D}({\bf r})$:
\begin{equation}
    \frac{\partial \widehat{J}_{ij}}{\partial {\bf r}_j} = -\frac{\partial \widehat{J}_{ij}}{\partial {\bf r}_i} = g_i g_j C_{dd} \left. \frac{\partial\widehat{D}}{\partial {\bf r}}  \right|_{{\bf r} = {\bf r}_{i,j,v}}
\end{equation}

Here the projection of the gradient $\partial \widehat{D}/\partial {\bf r}$ on an arbitrary direction ${\bf n} = (n_x, n_y, n_z)$ is 
\begin{equation}\label{Dderiv}
    \frac{\partial \widehat{D}}{\partial {\bf r}} = \frac{1}{|r|^4} \left[  
    3 \left(
    \begin{array}{ccc}
    n_x e_x & n_x e_y & n_x e_z \\
    n_y e_x & n_y e_y & n_y e_z \\
    n_z e_x & n_z e_y & n_z e_z
    \end{array}
    \right)
+
3 \left(
    \begin{array}{ccc}
    e_x n_x & e_x n_y & e_x n_z \\
    e_y n_x & e_y n_y & e_y n_z \\
    e_z n_x & e_z n_y & e_z n_z
    \end{array}
    \right)
    - 2 ({\bf e}\cdot {\bf n})\widehat{1} 
    \right] - 5\frac{\widehat{D}({\bf r}) ({\bf e} \cdot {\bf n}) }{|r|},
\end{equation}
where $({\bf e}{\bf n}) = {\bf n} {\bf r}/|r|$ and $\widehat{1}$ is a unit $3\times 3$ matrix. The Cartesian components of the gradient can be derived from Eq.~(\ref{Dderiv}) by considering ${\bf n}$ to be equal $(1,0,0)$, $(0,1,0)$ and $(0,0,1)$ respectively. 

It is noteworthy, that $\partial \widehat{D}/\partial {\bf r}$ decreases with the distance between spins faster then $\widehat{D}$ leading to a decreased importance of LRDD contributions to the magnon-phonon interaction. We include it only for 3D materials. In this case we introduce a vector of $3 \times 3$ matricies ${\bf D}_{{\bf k}}$:
\begin{equation}
{\bf D}'({\bf k}) = (D_x'({\bf k}), D_y'({\bf k}), D_z'({\bf k})) = \left( 
 \int_{|r|>R_{0}} \frac{\partial \widehat{D}}{\partial x}({\bf r}) e^{i{\bf k}{\bf r}} d{\bf r}, \, \int_{|r|>R_{0}} \frac{\partial \widehat{D}}{\partial y}({\bf r}) e^{i{\bf k}{\bf r}} d{\bf r}, \, \int_{|r|>R_{0}} \frac{\partial \widehat{D}}{\partial z}({\bf r}) e^{i{\bf k}{\bf r}} d{\bf r}  \right).
\end{equation}
Due to the spherical symmetry of the problem, ${\bf D}'({\bf k})$ for arbitrary direction of ${\bf k}$ can be calculated from the vector ${\bf D}'$ corresponding to a wavevector ${\bf k} = k_z {\bf e}_z$ in $z$-direction. 
\begin{equation}
 {\bf D}'({\bf k}) = T^T {\bf D}'(k_z {\bf e}_z) T,   
\end{equation}
where $T$ is a rotation matrix that brings ${\bf k}$ to $z$-direction. We derive explicit expression for ${\bf D}'(k_z {\bf e}_z)$:
\begin{equation}
 {D}'_x(k_z) = \frac{1}{R_0}\left(
 \begin{array}{ccc}
 0 & 0 & 3I_1 - 15 I_{1,2} \\
 0 & 0 & 0 \\
 3I_1 - 15 I_{1,2} & 0 & 0
 \end{array}\right), \quad 
 { D}'_y(k_z) = \frac{1}{R_0}\left(
 \begin{array}{ccc}
 0 & 0 & 0 \\
 0 & 0 & 3I_1 - 15 I_{1,2} \\
 0 & 3I_1 - 15 I_{1,2} & 0
 \end{array}\right)
\end{equation}
\begin{equation}
 { D}'_z(k_z) = \frac{1}{R_0}\left(
 \begin{array}{ccc}
 3I_1 - 15 I_{1,2} & 0 & 0 \\
 0 & 3I_1 - 15 I_{1,2} & 0 \\
 0 & 0 & 9I_1 - 15 I_{3}
 \end{array}\right)
\end{equation}
Here $I_1(x_0)$, $I_3(x_0)$ and $I_{1,2}(x_0)$ are analytical functions of $x_0 = k_z R_0$:
\begin{eqnarray}\label{I123}
I_1(x_0) =  -4\pi i \frac{x_0 \cos(x_0) +x_0^3 {\rm Ci}(x_0) - (1+x_0^2)\sin(x_0) }{3 x_0^2}
\nonumber    \\
I_3(x_0) = -4\pi i \frac{x_0 (x_0^2 - 6) \cos(x_0) + x_0^5 {\rm Ci}(x_0) -
(x_0^4 + 3 x_0^2 - 6)\sin(x_0)}{5x_0^4}
\\
I_{1,2}(x_0) = \frac{I_1(x_0) -  I_3(x_0)}{2}
\nonumber
\end{eqnarray}

The modifications of the exchange and dipole–dipole interactions described in this section are incorporated into the magnon–phonon interaction Hamiltonian through the Holstein–Primakoff transformation [Eq.~(\ref{HP1})] and the standard second quantization of phonons [Eq.~(\ref{phon})].

\subsection{Evaluation of Interaction Matrix Elements}

Here we outline the procedure for deriving the four-magnon matrix elements from the spin Hamiltonian, as well as the two-magnon–one-phonon matrix elements arising from its modification by atomic displacements. We focus on MC processes; the treatment of magnon-nonconserving MNC processes is similar.

The derivation of the four-magnon matrix elements begins by expressing each term of the spin Hamiltonian in its Holstein–Primakoff representation and retaining the corresponding quartic magnon contributions. To illustrate the procedure, we consider a generic interaction term ${\bf S}_1 \widehat{J} {\bf S}_2$, which describes the coupling between two arbitrary spins, labeled $1$ and $2$, via a $3 \times 3$
exchange interaction matrix $\widehat{J}$. The resulting four-magnon terms are
\begin{multline}\label{HP4o}
{\bf S}_1 \widehat{J} {\bf S}_2 \rightarrow J_{zz}a_1^+ a_2^+ a_1 a_2 \\
-\frac{1}{8}\sqrt{\frac{S_2}{S_1}}(J_{xx} + J_{yy} + i(J_{xy}-J_{yx})) a_1^+ a_2^+ a_1 a_1 
-\frac{1}{8}\sqrt{\frac{S_1}{S_2}}(J_{xx} + J_{yy} + i(J_{xy}-J_{yx})) a_2^+ a_2^+ a_1 a_2 \\
-\frac{1}{8}\sqrt{\frac{S_2}{S_1}}(J_{xx} + J_{yy} - i(J_{xy}-J_{yx})) a_1^+ a_1^+ a_1 a_2 
-\frac{1}{8}\sqrt{\frac{S_1}{S_2}}(J_{xx} + J_{yy} - i(J_{xy}-J_{yx})) a_1^+ a_2^+ a_2 a_2  \\
-\frac{1}{8}\sqrt{\frac{S_2}{S_1}}(J_{xx} - J_{yy} - i(J_{xy} + J_{yx})) a_1^+ a_1 a_1 a_2 
-\frac{1}{8}\sqrt{\frac{S_1}{S_2}}(J_{xx} - J_{yy} - i(J_{xy}+ J_{yx})) a_2^+ a_1 a_2 a_2 \\
-\frac{1}{8}\sqrt{\frac{S_2}{S_1}}(J_{xx} - J_{yy} + i(J_{xy}+ J_{yx})) a_1^+ a_1^+ a_2^+ a_1 
-\frac{1}{8}\sqrt{\frac{S_1}{S_2}}(J_{xx} - J_{yy} + i(J_{xy}+ J_{yx})) a_1^+ a_2^+ a_2^+ a_2  
\end{multline}
Here the arrow $\rightarrow$ denotes the Holstein–Primakoff transformation [Eq.~(\ref{HP1})], followed by the selection of fourth-order terms in the magnon creation and annihilation operators associated with spins $1$ and $2$, namely  $a_1^+$, $a_2^+$, $a_1$ and $a_2$. 

We focus on the contribution of expression~(\ref{HP4o}) to the Hamiltonian term describing the annihilation of an acoustic magnon with wave vector ${\bf k}_0 \rightarrow 0$ and a second magnon with wave vector ${\bf k}_1$ and branch $\alpha_1$, accompanied by the creation of two magnons with wave vectors
${\bf k}_2$ and ${\bf k}_3$ and branches ${\alpha}_2$ and ${\alpha}_3$, respectively. Acoustic magnon branch corresponds to $\alpha =1$ in this section.

The corresponding contribution arises from the operator combination
\begin{equation}\label{s4MC}
s_{\alpha_2}^+({\bf k}_2) s_{\alpha_3}({\bf k}_3)^+ s_{\alpha_1}({\bf k}_1) s_{1}(0),
\end{equation}
where $s^+_{\alpha}({\bf k})$ and $s_{\alpha}({\bf k})$ denote the creation and annihilation operators of the diagonalized magnon mode $\alpha$ with a wave vector ${\bf k}$. Because magnon creation and annihilation operators are mixed during the diagonalization procedure, each of the original operators
$a_1^+$, $a_2^+$, $a_1$ and $a_2$ can contribute to any of the final operators
$s_{\alpha_2}^+({\bf k}_2)$, $ s_{\alpha_3}^+({\bf k}_3)$, $s_{\alpha_1}({\bf k}_1)$ and $ s_{1}(0)$. All such mappings are included in the evaluation of the matrix elements within our algorithm. Below, we explicitly present the expressions corresponding to a single representative mapping.
\begin{equation}
a_1^+ \rightarrow s_{\alpha_2}^+({\bf k}_2), \quad a_2^+ \rightarrow s_{\alpha_3}^+({\bf k}_3), \quad a_1 \rightarrow s_{\alpha_1}({\bf k}_1), \quad a_2 \rightarrow s_{1}(0).
\end{equation}
This mapping yields the following contribution to the Hamiltonian describing the four-magnon processes:
\begin{multline}\label{mapRes1}
\frac{J_{zz}}{N^2} \exp\left( i({\bf k}_2{\bf r}_1 + {\bf k}_3{\bf r}_2 - {\bf k}_1{\bf r}_1)\right) \\
\times
G_{m+i_1, m+\alpha_2}(-{\bf k}_2) G_{m+i_2, m+\alpha_3}(-{\bf k}_3)
G_{i_1, m-\alpha_1}(-{\bf k}_1)G_{i_2, m-1}(0)
s_{\alpha_2}^+({\bf k_2})s_{\alpha_3}^+({\bf k_3}) s_{\alpha_1}({\bf k_1})s_{1}(0).
\end{multline}
Here $N$ denotes the number of unit cells in the material, and ${\bf r}_1$ and ${\bf r}_2$ are the positions of spins $1$ and $2$, respectively. Owing to momentum conservation, ${\bf k}_2 + {\bf k}_3 - {\bf k}_1 = 0$, only the relative positions of the spins are important for the result.  The proportionality of Eq.~(\ref{mapRes1}) to $J_{zz}$ reflects the fact that the simultaneous presence of all four operators $a_1^+$, $a_2^+$, $a_1$ and $a_2$
occurs exclusively in the first term of Eq.~(\ref{HP4o}). The matrix $\widehat{G}({\bf k})$ is the standard diagonalization matrix of linear spin-wave theory, defined as follows:
\begin{equation}\label{G}
\left( 
\begin{array}{c}
a_{1}({\bf k}) \\ ...\\ a_{m}({\bf k}) \\
a_{1}^+(-{\bf k}) \\ ... \\ a_{m}^+(-{\bf k})
\end{array}
\right) = \widehat{G}({\bf k})
\left(
\begin{array}{c}
s_m({\bf k}) \\ ...\\ s_1({\bf k}) \\
s_1^+(-{\bf k}) \\ ... \\ s_m^+(-{\bf k})
\end{array}
\right).
\end{equation}
Here index $1,...m$ in l.h.s. of the equation enumerates the $m$ different magnetic atoms in a unit cell (we also call them types of magnetic atoms). In r.h.s. the same index enumerates magnon branches. $a_{i}({\bf k})$ is the Fourier transform of the annihilation operators $a_{i,v}$  of all the atoms of the type $i$ belonging to the different unit cells $v$: 
\begin{equation}
a_{i}({\bf k}) = \frac{1}{\sqrt{N}}\sum_{v} a_{(i,v)} \exp(-i{\bf k} \cdot {\bf r}_{(i,v)}).
\end{equation}
Here ${\bf r}_{i, v}$ is the position of the magnetic atom of type $i$ and belonging to the unit cell $v$.

The contribution of the LRDD interaction to the four-magnon matrix elements is evaluated in an analogous manner. The only difference is that the Holstein–Primakoff decomposition of the Hamiltonian [Eq.~(\ref{HLRDD})] is performed after the Fourier transformation, resulting in momentum-space magnon operators $a_{i}({\bf k})$ and $a_{i}^+({\bf k})$ rather than on-site magnon creation and annihilation operators.
Finally, we note that a double counting arises because the two emitted magnons, $({\bf k}_2, \alpha_2)$ and $({\bf k}_3, \alpha_3)$ can be interchanged. This is accounted for by multiplying the squared matrix element $|W_{if}|^2$ by a factor of $1/2$ in the final expression~(\ref{B-4m-MC}). 

The general strategy for deriving the magnon–phonon matrix elements is similar. In this case, however, the matrix elements originate from the terms 
\begin{equation}
{\bf S}_1 \frac{\partial \widehat{J}}{\partial r_k} {\bf S}_2 u_k
\end{equation}
in the general spin Hamiltonian [Eq.~(1) of the main text]. Here, the index $k$ labels both the displaced atoms and the Cartesian coordinates. The atomic displacement $u_k$ is expressed in terms of phonon creation and annihilation operators using Eq.~(\ref{phon}), while the spin-dependent part of the Hamiltonian is reduced to second-order terms in the magnon creation and annihilation operators with Eq.~(\ref{HP1}). The second procedure yields
\begin{multline}\label{HP4o-2}
{\bf S}_1 \widehat{J}' {\bf S}_2 \rightarrow -J_{zz}' S_2 a_1^+ a_1 -J_{zz}' S_1 a_2^+ a_2 \\
+ a_{1} a_{2}\frac{\sqrt{S_1 S_2}}{2} \left(J_{xx}' - J_{yy}' - i(J_{xy}' + J_{yx}')\right) 
+  a_{1}a_{2}^+\frac{\sqrt{S_1 S_2}}{2} \left(J_{xx}' + J_{yy}' + i(J_{xy}' - J_{yx}')\right) \\
+  a_{1}^+a_{2}\frac{\sqrt{S_1 S_2}}{2} \left(J_{xx}' + J_{yy}' - i(J_{xy}' - J_{yx}')\right) 
+  a_{1}^+a_{2}^+\frac{\sqrt{S_1 S_2}}{2} \left(J_{xx}' - J_{yy}' + i(J_{xy}' + J_{yx}')\right),
\end{multline}
where $\widehat{J}' \equiv \partial \widehat{J}/\partial r_k$. The following transformation from the on-site magnon operators $a_1^+$ and $a_1$ to the diagonilized magnon operators $s^+_{\alpha}({\bf k})$ and $s_{\alpha}({\bf k})$ is performed similarly to the 4-magnon case. 

\subsection{Algorithmic Workflow}

\begin{figure*}
 	\centering
 	\includegraphics[width=7.0in]{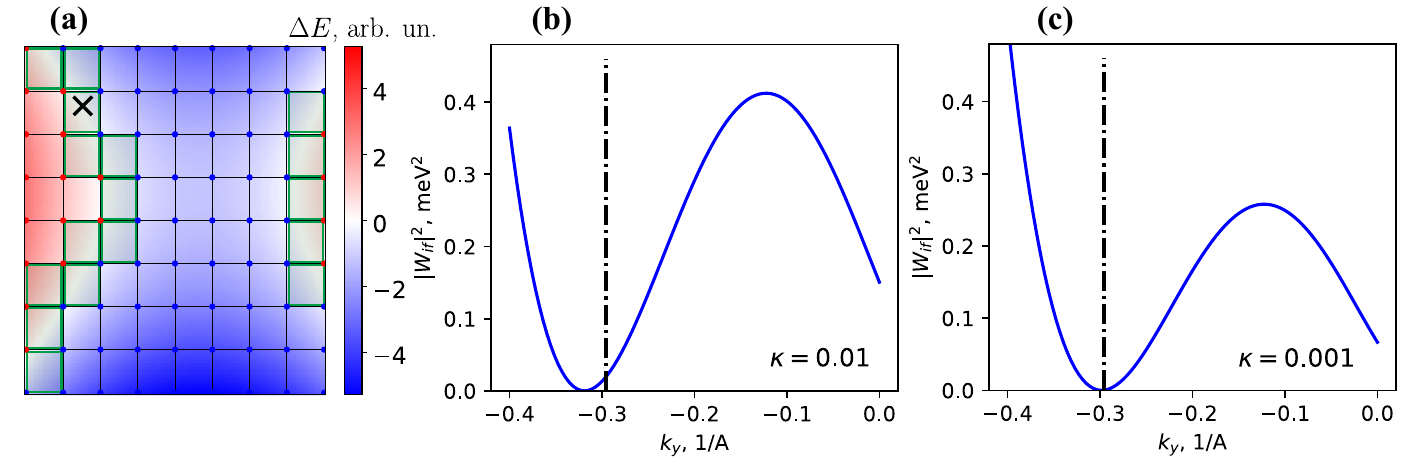}
 	\caption{ \label{fig:SI-algol} (a) Illustration of the algorithm workflow. The energy differences
    $\Delta E = E_i-E_f$ evaluated at each node of the ${\bf k}$-grid are shown as blue points when negative and red points when positive. Green squares indicate grid cells in which \textit{scattering events} are expected. Black crosses mark individual \textit{scattering events} identified within each such cell (shown for one representative cell only).(b,c) Dependence of the calculated squared matrix element $|W_{if}|^2$ on $k_{y}$ -- the $y$-component of the wave vector of the first emitted magnon, for magnon–magnon damping in a two-dimensional toy model. The black dot-dashed line indicates the value of $k_{y}$ corresponding to the energy conservation. Panel (b) corresponds to the model parameters used in the main text, while panel (c) shows the same model with
    $\kappa = 10^{-3}$. 
 	   } 
\end{figure*}

We begin the damping simulations by introducing a Monkhorst–Pack $k$-space grid (Grid 1). When only low-${\bf k}$ quasiparticle contributions are expected to be significant, a cutoff parameter $0<r_m<1$
may be applied to restrict the grid to the central region around the $\Gamma$ point. This grid represents the phonon wave vector in magnon–phonon (2m–1ph) processes and the wave vector ${\bf k}_1$ of the first absorbed magnon in four-magnon (4m) processes. In the latter case, for each value of ${\bf k}_1$, 
an additional grid (Grid 2) is constructed to sample the wave vector ${\bf k}_2$ of the second magnon. The wave vector of the remaining interacting quasiparticle is then determined from momentum conservation.

For each point of Grid 1 (for 2m–1ph processes) or Grid 2 (for 4m processes), we evaluate the energy difference $\Delta E = E_i - E_f$ entering the Boltzmann equation. Within a given ${\bf k}$-grid cell, changes in the sign of $\Delta E$ among different cell corners indicate the presence of \textit{scattering events}, i.e., sets of quasiparticle wave vectors that satisfy both energy and momentum conservation. This procedure is illustrated in Fig.~\ref{fig:SI-algol}(a), where grid cells containing \textit{scattering events} are marked by green squares. For each such cell, the code identifies a single representative scattering event: a set of wave vectors for which $\Delta E = 0$ is satisfied with high numerical precision. One such event is indicated by a black cross in a representative cell in Fig.~\ref{fig:SI-algol}(a). The matrix element $W_{if}$ (see Eq.~(\ref{bolt0})) and the occupation numbers of the interacting quasiparticles are evaluated at the wave vectors corresponding to this scattering event and assumed to be constant within the cell. Consequently, the $\bf k$-space the cell reduces to the integration of the Dirac delta function enforcing energy conservation. This integral is performed using a bilinear (in 2D) or trilinear (in 3D) interpolation of $\Delta E$ within the cell, combined with the marching cubes method~\cite{marching}.
The total integral over the ${\bf k}$-space grid is then obtained as a sum of contributions from all such cells, or equivalently, from all identified \textit{scattering events}. For 2m–1ph processes, this procedure yields the magnon decay rate directly. For 4m processes, an additional summation over Grid 1 is required, corresponding to the integration over the wave vector ${\bf k}_1$.

Our procedure ensures that matrix elements are evaluated only for wave vectors that satisfy the energy and momentum conservation laws. This is essential because the squared matrix element $|W_{if}|^2$ typically exhibits a pronounced minimum in the vicinity of these conservation points. This minimum becomes particularly pronounced when spin–orbit and dipole–dipole interactions are weak. This behavior is illustrated in Fig.~\ref{fig:SI-algol}(b) and (c), where we consider magnon–magnon damping in a two-dimensional toy model. Figure~\ref{fig:SI-algol}(b) corresponds to the model parameters used in the main text, while Fig.~\ref{fig:SI-algol}(c) shows the same model with a reduced anisotropy parameter $\kappa = 10^{-3}$. In both cases, we fix a single \textit{scattering event} and treat the $y$-component of the wave vector ${\bf k}_2$ of the first emitted magnon as a free parameter $k_y$. The wave vector of the second emitted magnon is chosen as ${\bf k}_3 = {\bf k}_1 - {\bf k}_2$ to ensure momentum conservation; however, energy conservation is satisfied only for a single “correct” value of $k_y$, indicated by the dash-dotted line in Figs.~\ref{fig:SI-algol}(b) and (c).
Despite this, we compute the squared matrix element $|W_{if}|^2$ for all values of $k_y$ and display the results in the figure. Notably, $|W_{if}|^2$ exhibits a clear minimum —- and in fact tends toward zero—at a value of $k_y$ very close to the energy-conserving one. This behavior becomes increasingly pronounced in Fig.~\ref{fig:SI-algol}(c), corresponding to the small anisotropy parameter $\kappa = 10^{-3}$. 

Typically, the most time-consuming part of the algorithm is the identification of \textit{scattering events}, followed by the evaluation of the corresponding matrix elements. However, within the Boltzmann framework, both the quasiparticle energies and the interaction Hamiltonian are independent of temperature. As a result, the damping can be efficiently computed across the entire temperature range, with the \textit{scattering events} needed to be determined only once.

\section{Detailed \textit{ab initio} Results for YIG and Monolayer $\rm CrSBr$}

Although magnons in insulators can be viewed as collective electronic excitations~\cite{Green0, AliPh,Ali2}, we base our modeling on spin Hamiltonians derived from \textit{ab initio} calculations.

\textit{Ab initio} simulations of YIG and monolayer CrSBr were performed using VASP~\cite{VASP} and SIESTA~\cite{SIESTA} codes. Tight-binding Hamiltonians were constructed with wannier90~\cite{wannier90}, and spin Hamiltonians were derived via perturbation theory using TB2J~\cite{TB2J} software. The resulting magnon dispersions, calculated with RAD-tools package~\cite{RadTools}, agree well with previously reported results~\cite{YIGJex,CrSBrStrain}. Phonon calculations were carried out using the finite-displacement method implemented in phonopy~\cite{phonopy}. Spin–orbit coupling was included for CrSBr, whereas YIG was treated in the non-relativistic limit. Consequently, damping in YIG arises solely from dipole–dipole interactions added at the post-processing stage.
The damping coefficient $\alpha$ is computed numerically using our MagnoFallas package~\cite{MagnoFallas}.

\subsection{Damping in a Monolayer of CrSBr}

\begin{figure}
 	\centering
 	\includegraphics[width=5.0in]{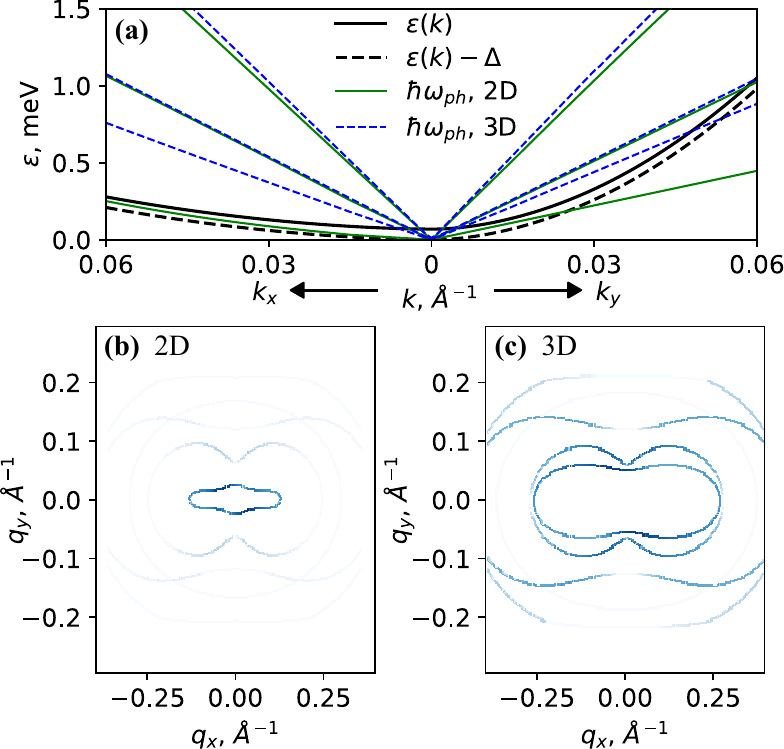}
 	\caption{ \label{fig:phoncon}
    (a) Low-energy magnon dispersion (black solid line) compared with the phonon dispersions obtained from \textit{ab initio} calculations for a CrSBr monolayer (green solid line) and bulk CrSBr (blue dashed line). The black dashed line indicates the phonon energy that must be absorbed in the MC process. (b) Contributions of phonons with different wavevectors to the MC magnon–phonon damping in the case of two-dimensional phonons. (c) Corresponding contributions for three-dimensional phonons. } 
\end{figure}

In this section we describe in detail the simulation results for ${\rm CrSBr}$. Because the layers of vdW materials can be obtained either as free-hanging films of be put on a substrate, we introduce two distinct models for the magnon-phonon damping: the models of "2D phonons" and the model of "3D phonons". In the first (2D) version we calculate the phonons from the monolayer of ${\rm CrSBr}$. In the second version (3D phonons) we derive phonons from bulk ${\rm CrSBr}$, nevertheless we still consider a 2D spin Hamiltonian. The magnon-phonon damping in the two cases appears to be $\sim 10$ times different as shown in the main text.

The pronounced difference between the two cases can be understood from the results shown in Fig.~\ref{fig:phoncon}. Figure~\ref{fig:phoncon}(a) compares the magnon dispersion with the dispersions of 2D and 3D phonons. In the 2D case, the phonon branch corresponding to the acoustic mode with atomic displacements perpendicular to the monolayer exhibits significantly lower energies, consistent with the presence of flexural phonons~\cite{Landau7}. As a result, crossings between the phonon and magnon spectra occur at much lower energies. Figures~\ref{fig:phoncon}(b) and (c) show the contributions of phonons with different wavevectors to the damping for 2D and 3D phonons, respectively. In the 2D case, the damping is dominated by low-wavevector flexural phonons, which have low energies and correspondingly large occupation numbers. In contrast, for 3D phonons the damping is governed by phonons with higher energies and smaller occupation numbers, resulting in an approximately 10-fold reduction of the damping parameter $\alpha$.

\begin{figure*}
 	\centering
 	\includegraphics[width=5.0in]{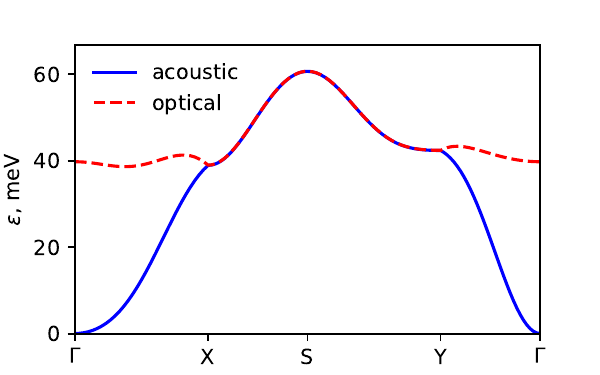}
 	\caption{ \label{fig:SI-CrSBr-disp}  Magnon dispersion in a monolayer of CrSBr.
 	   } 
\end{figure*}

\begin{figure*}
 	\centering
 	\includegraphics[width=7.0in]{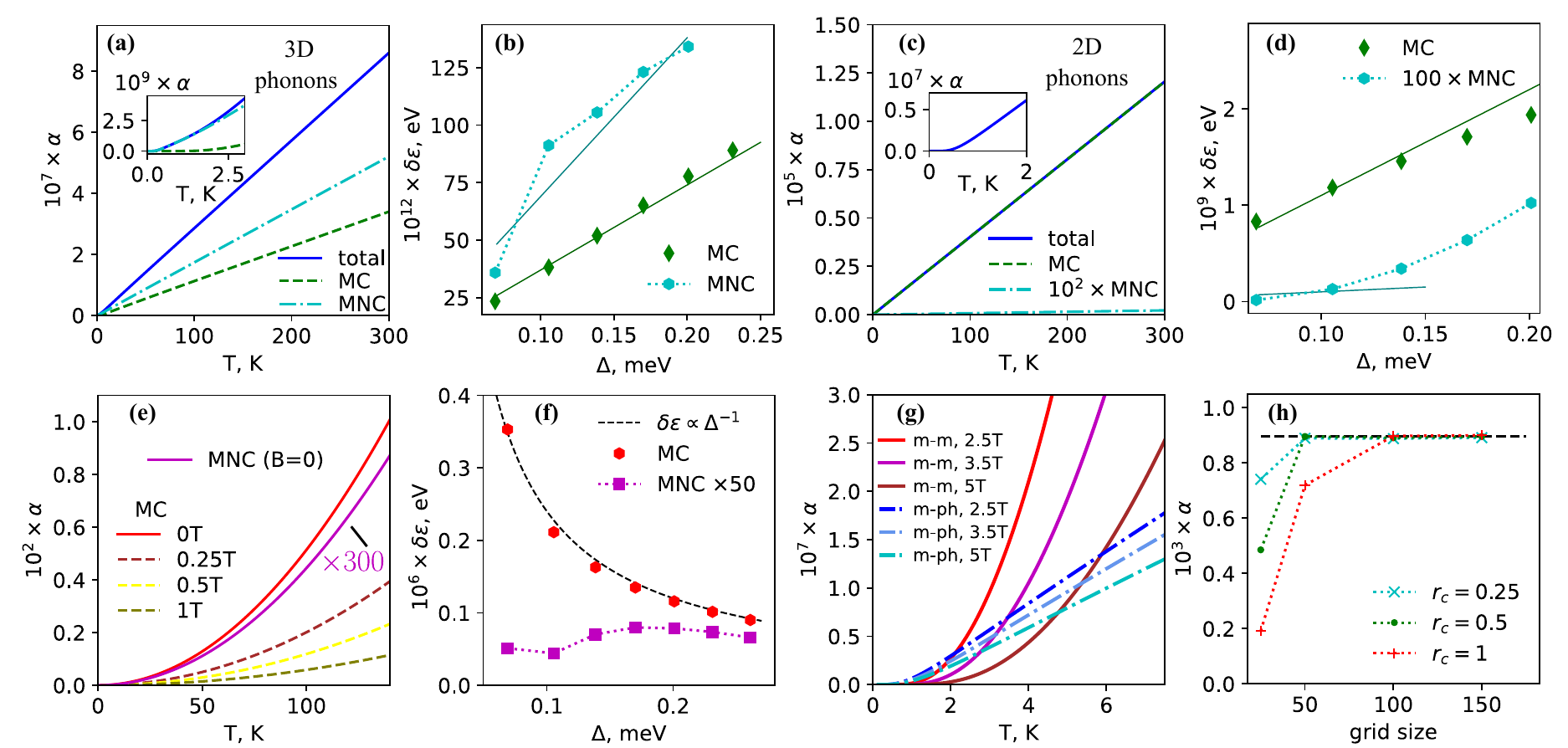}
 	\caption{ \label{fig:SI-CrSBr}  The results of the numeric simulations of damping in ${\rm CrSBr}$. (a) the contributions of MC and MNC processes to the magnon-phonon damping mechanism in case of the phonons derived from bulk CrSBr. Inset shows low temperatures $T\sim 1{\, K}$. (b) FMR linewidth dependency on $\Delta$ for the same processes compared with Gilpert damping. (c) The contributions of MC and MNC processes to the magnon-phonon damping mechanism in case of the phonons derived from a CrSBr monolayer. (d) Comparison of the FMR linewidth with the Gilbert law for these processes. (e) damping constant calculated from MC magnon-magnon scattering  at diferent magnetic field compared to the damping due to MNC magnon-magnon processes at zero magnetic field. (f) FMR linewidth due to MC and MNC magnon-magnon processes compared with the non-Gilbert law $\delta \varepsilon \propto \Delta^{-1}$. (g) Damping constant due to magnon-magnon and magnon-phonon processes calculated at very high magnetic fields and low temperatures. (h) magnon-magnon damping calculated with different $k$-grid for zero external magnetic field and $T=40\,{\rm K}$.
 	   } 
\end{figure*}

Figure~\ref{fig:SI-CrSBr-disp} shows the calculated magnon dispersion in a monolayer of CrSBr.
It is similar to the one reported in previous studies \cite{CrSBrwaves,CrSBrStrain}.
Figure~\ref{fig:SI-CrSBr} presents detailed results of our simulations of damping in a CrSBr monolayer. Figure~\ref{fig:SI-CrSBr}(a) shows the contributions of MC and MNC magnon–phonon processes in the simulation with the phonons derived from a 3D CrSBr (which qualitatively represents a monolayer on a substrate). The two contributions are comparable in magnitude, similar to the behavior observed in YIG. Figure~\ref{fig:SI-CrSBr}(b) demonstrates that both contributions exhibit approximately Gilbert-like behavior, although deviations from Gilbert law are significantly stronger for MNC processes. 

Figures~\ref{fig:SI-CrSBr}(c) and (d) present the corresponding results when the phonons are derived from a monolayer itself (representing a free-hanging monolayer in vacuum). In this case, magnon–phonon damping is dominated by MC processes, while the MNC contribution becomes non-Gilbert already at relatively weak magnetic fields, as illustrated in Fig.~\ref{fig:SI-CrSBr}(d).

Figure~\ref{fig:SI-CrSBr}(e) shows the MC contribution to the magnon–magnon damping calculated for magnetic fields ranging from $0$ to $1,\mathrm{T}$. The damping is already strongly suppressed at $0.25,\mathrm{T}$ and decreases further with increasing field. Nevertheless, even at $1,\mathrm{T}$ the MC magnon–magnon contribution remains significantly larger than the MNC contribution at zero field. Figure~\ref{fig:SI-CrSBr}(f) illustrates the dependence of the FMR linewidth associated with magnon–magnon scattering on the magnon gap $\Delta$. The MC contribution decreases with increasing $\Delta$, in agreement with the toy model, whereas the MNC contribution is nearly independent of the magnetic field.

\begin{figure*}
 	\centering
 	\includegraphics[width=7.0in]{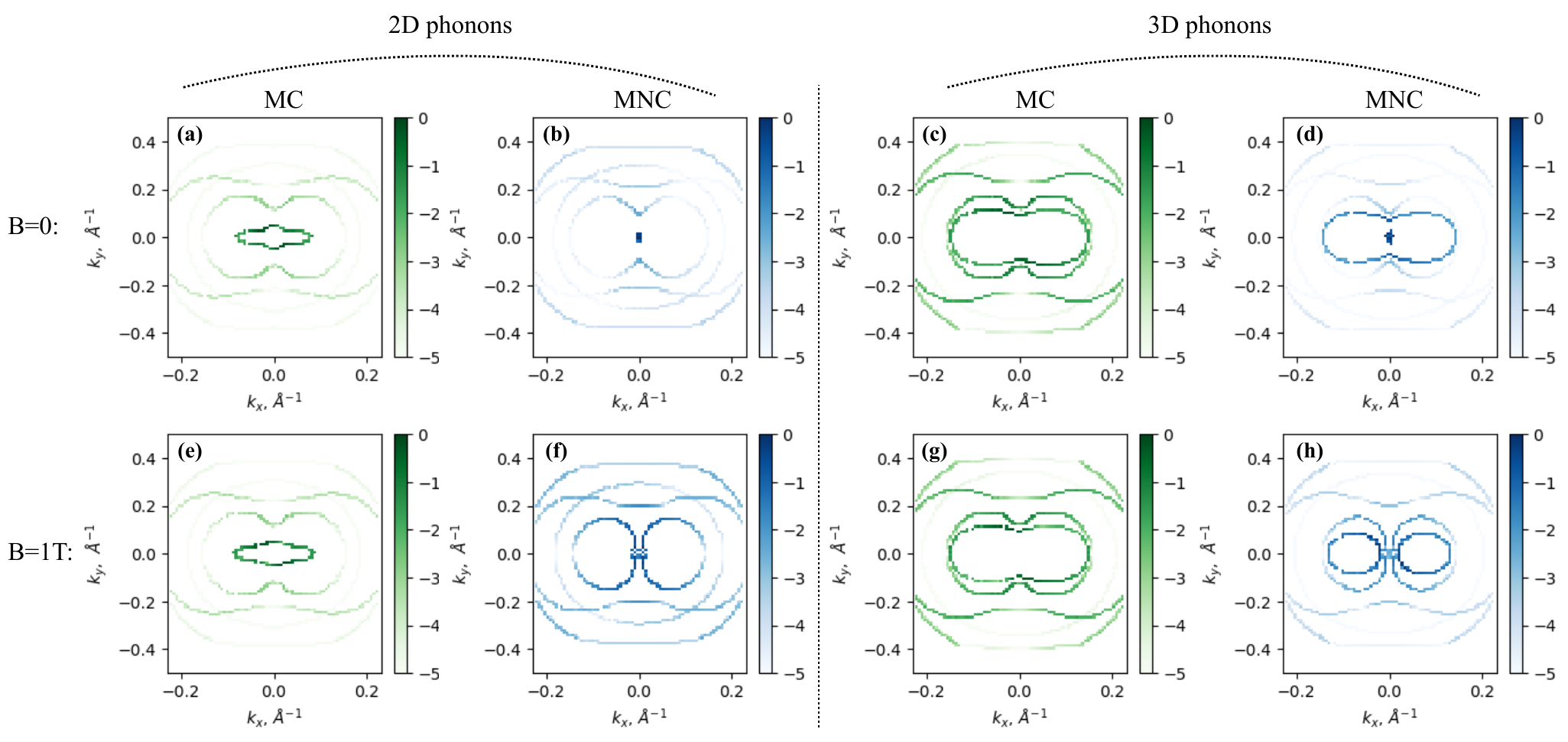}
 	\caption{ \label{fig:SI-CrSBr-phonmap} 
    The contribution $p({\bf k})$ of different phonon wavevectors ${\bf k}$ to the magnon-phonon damping in ${\rm CrSBr}$ in zero magnetic field (a-d) and at $B=2\,{\rm T}$ (e-h). All the panels show ${\rm log}_{10}(p({\bf k})/p_{max})$ where $p_{max} = {\rm max}(p({\bf k}))$. Panels (a,c,e,g) correspond to MC process and panels  (b,d,f,h)  show MNC processes. Panels (a,b,e,f) correspond to 2D phonons model (free-hanging layer of ${\rm CrSBr}$) and  panels (c,d,g,h) to the 3D model ({\rm CrSBr} on a substrate). 
 	   } 
\end{figure*}

Figure~\ref{fig:SI-CrSBr}(h) illustrates the dependence of the calculated damping on the choice of the $k$-grid. We show the magnon--magnon damping calculated at zero external magnetic field and at $T=40,\mathrm{K}$. Three integration regions are considered: the entire first Brillouin zone ($r_c=1$), a region centered at the $\Gamma$ point covering one quarter of the Brillouin zone ($r_c=0.5$), and a region covering only one sixteenth of the Brillouin zone ($r_c=0.25$). For each case, four grid sizes ranging from $25\times25$ to $150\times150$ were employed. The results indicate that, without a cutoff ($r_c=1$), relatively dense grids are required to achieve convergence. Introducing a cutoff allows the use of smaller grids by restricting the calculation to the low-$k$ magnons that provide the dominant contribution to the damping. Nevertheless, for sufficiently dense grids, the results obtained for different values of $r_c$ agree to within $1\%$, demonstrating that the calculated damping is robust with respect to the choice of $k$-grid.

Under all conditions considered so far, the magnon–magnon contribution to the damping is substantially larger than the magnon–phonon contribution. However, the magnon–phonon damping scales linearly with temperature, whereas the magnon–magnon contribution scales quadratically and can be efficiently suppressed by an external magnetic field. As a result, magnon–phonon interactions are expected to dominate the intrinsic damping in a CrSBr monolayer at sufficiently low temperatures and high magnetic fields. To test this expectation, we calculated the damping for magnetic fields of $2.5$, $3.5$, and $5\,\mathrm{T}$ and for temperatures below $7\,\mathrm{K}$. The results, shown in Fig.~\ref{fig:SI-CrSBr}(g), indicate that the magnon–phonon contribution becomes dominant at magnetic fields of approximately $3.5\,\mathrm{T}$ and temperatures below $\sim 4\,\mathrm{K}$.

While the toy model predicts Gilbert-like damping for the magnon–phonon mechanism, deviations from the Gilbert behavior are observed in the calculated damping, particularly for MNC magnon–phonon processes. Equation (9) in the main text indicates that the damping is strictly Gilbert-like when the energies of the participating quasiparticles, as well as the matrix elements $W_{if}$, are independent of the external magnetic field. This condition is satisfied in the toy model when only magnon-number-conserving (MC) processes are considered and the magnetic field contributes merely as an additive shift to the magnon energies, identical for all magnon modes. In real materials, however, MNC processes also contribute to magnon–phonon damping. These processes satisfy the energy conservation condition [$\Delta + \varepsilon({\bf q}) = \hbar \omega_{\bf q}$
], where $\varepsilon({\bf q})$ and $\hbar \omega_{\bf q}$ are the magnon and phonon energies, respectively, associated with the same wavevector ({\bf q}). When an external magnetic field increases the magnon energies ($\Delta$ and $\varepsilon({\bf q})$), the phonon energy participating in the process must also change to maintain energy conservation. As a consequence, the wavevector ${\bf q}$ may be modified. Therefore, the Gilbert law is expected to hold only approximately and primarily in the limit of weak magnetic fields. Furthermore, even for MC processes, the presence of spin–orbit and dipole–dipole interactions can give rise to a complex magnetic-field dependence of the magnon spectrum, rather than a simple additive energy shift. Such effects can also produce deviations from Gilbert-like damping. Nevertheless, these deviations are expected to be less pronounced than those associated with MNC processes.

To illustrate this effect, Fig.~\ref{fig:SI-CrSBr-phonmap} shows the relative contribution ($p({\bf k})$) of phonons with wavevector ${\bf k}$ to the magnon–phonon damping in ${\rm CrSBr}$, calculated both at zero magnetic field and at $B=1\,{\rm T}$. In all the cases ${\rm log}_{10}(p({\bf k})/p_{max})$ is shown where $p_{max}$ is the maximal contribution. Results are presented for both MC and MNC processes, as well as for the two phonon models considered (2D and 3D phonons). In all cases, the phonon modes contributing to the MNC processes change substantially when the magnetic field is increased from $0$ to $1\,{\rm T}$, indicating that the corresponding damping is already expected to deviate from ideal Gilbert behavior at this field strength. Smaller but still noticeable changes are also observed for the MC processes. These modifications are likely responsible for the weak deviations from Gilbert-like behavior found in the MC magnon–phonon damping.

\subsection{Damping in Bulk YIG}

\begin{figure*}
 	\centering
 	\includegraphics[width=7.0in]{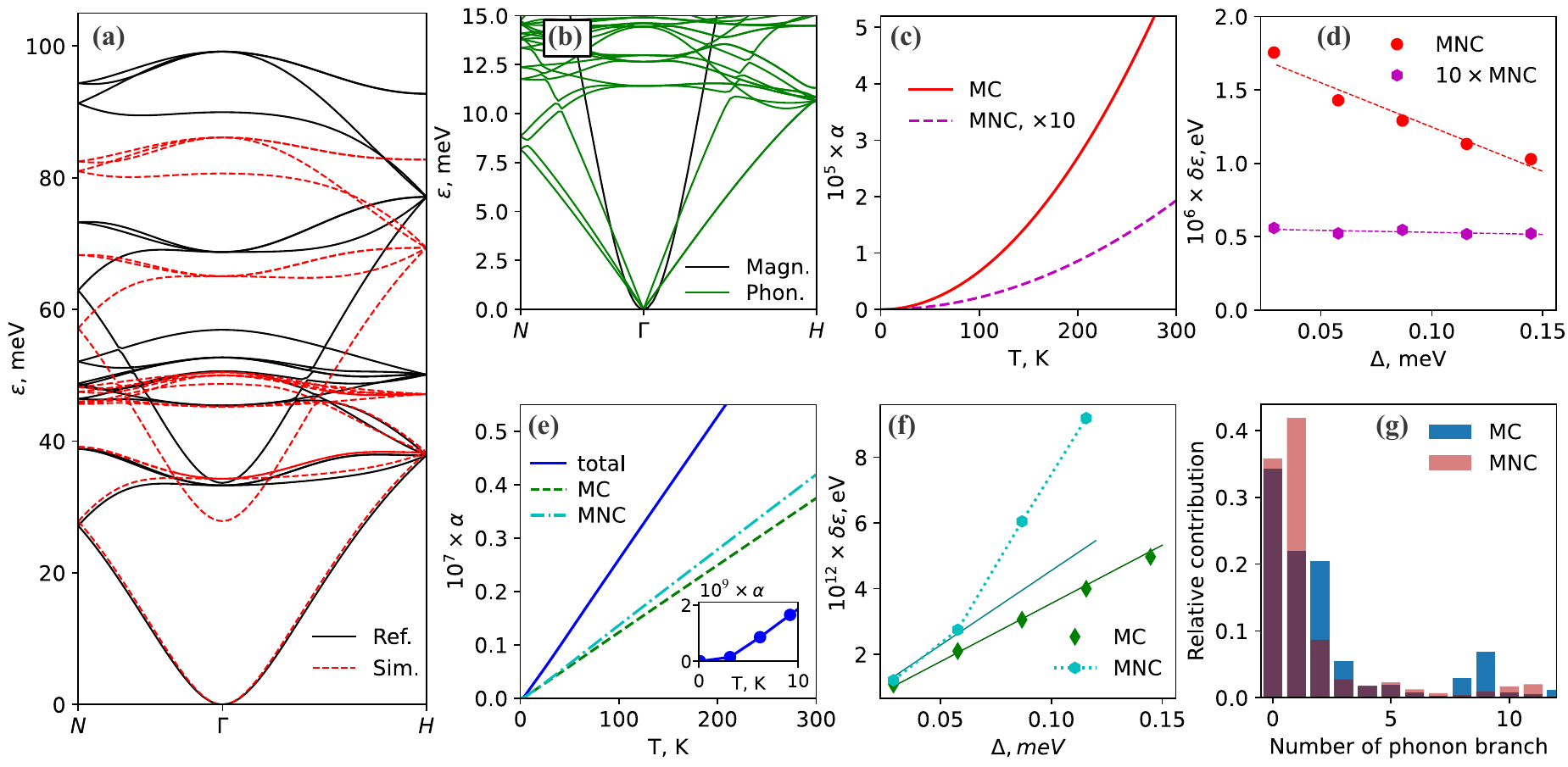}
 	\caption{ \label{fig:SI-YIG} 
    Results of the numerical simulations of damping in YIG.
    (a) Magnon dispersion calculated using the exchange interactions from Ref.~\cite{YIGJex} (black solid line) compared with our SIESTA-based calculations (red dashed line).
    (b) Magnon dispersion (black) and phonon dispersion (green).
     (c) Contributions of magnon-conserving MC and MNC processes to the magnon–magnon damping in YIG as a function of temperature.
     (d) FMR linewidth arising from MC and MNC magnon–magnon processes as a function of the magnon gap $\Delta$.
     (e) Contributions of MC and MNC magnon–phonon processes to the magnon–phonon damping.
     (f) FMR linewidth due to MC and MNC magnon–phonon processes compared with the Gilbert law.
    (g) Relative contributions of different phonon branches to the magnon–phonon damping at $T=300,\mathrm{K}$, calculated for both MC and MNC processes.
 	   } 
\end{figure*}

The Heisenberg component of the spin Hamiltonian in YIG has been investigated \textit{ab initio} in several works and extensively compared with experimental data. In this study, we adopt the Heisenberg exchange parameters from Ref.~\cite{YIGJex}. The dipole–dipole interactions are added to the Hamiltonian using the atomic coordinates of the crystal and are essential for capturing magnon damping.

The phonon dispersion and the magnon–phonon interaction were obtained from additional calculations based on the finite-displacement method, employing the SIESTA~\cite{SIESTA}, phonopy~\cite{phonopy}, wannier90~\cite{wannier90}, and TB2J~\cite{TB2J} software packages. All \textit{ab initio} calculations were performed in the nonrelativistic limit, yielding only Heisenberg exchange interactions; the dipole–dipole interaction was incorporated \textit{a posteriori}. Modulations of the Heisenberg exchange induced by atomic displacements do not contribute to damping when magnon states are also computed in the nonrelativistic limit. However, when magnon states include dipole–dipole interactions, such contributions become symmetry-allowed and were therefore included in the simulations. Nonetheless, we find that their contribution is negligible compared to the effect of the displacement-induced modulation of the dipole–dipole interaction.

The detailed simulation results are presented in Fig.~\ref{fig:SI-YIG}. Figure~\ref{fig:SI-YIG}(a) compares the magnon dispersion obtained using the exchange parameters of Ref.~\cite{YIGJex} with our SIESTA-based calculations. Although discrepancies appear at higher magnon energies, the low-energy portion of the dispersion -— most relevant for damping —- is reproduced accurately. Figure~\ref{fig:SI-YIG}(b) shows the magnon and phonon dispersions in YIG. Up to energies of approximately $\varepsilon \sim 7,\mathrm{meV}$, the phonon spectrum consists of three acoustic branches, while a large number of optical modes appear at higher energies. The two lowest phonon branches intersect the magnon dispersion at relatively low energies, $\varepsilon \sim 1,\mathrm{meV}$.

Figures~\ref{fig:SI-YIG}(c) and (d) demonstrate that the magnon–magnon contribution to the damping is dominated by MC processes, whereas the contribution from MNC processes is approximately 30 times smaller. The damping arising from both MC and MNC magnon–magnon processes is non-Gilbert in nature: the linewidth associated with MC processes decreases linearly with the ${\bf k}=0$ acoustic magnon gap $\Delta_0$, in agreement with our toy model, while the linewidth due to MNC processes is nearly independent of $\Delta_0$.

The contributions of MC and magnon-nonconserving MNC processes to magnon–phonon damping are shown to be comparable in Fig.~\ref{fig:SI-YIG}(e). The inset of this figure displays the temperature dependence of the total magnon–phonon damping at low temperatures, $T<10\,\mathrm{K}$. The damping exhibits a linear temperature dependence starting from $T \sim 3\,\mathrm{K}$ and is strongly suppressed at very low temperatures, in agreement with the toy model.

Figure~\ref{fig:SI-YIG}(f) compares the ferromagnetic resonance (FMR) linewidth arising from MC and MNC magnon–phonon processes with the Gilbert form. The MC contribution follows the Gilbert law over the entire magnetic-field range considered ($0.25$–$1.25,\mathrm{T}$), whereas the MNC contribution begins to deviate from Gilbert behavior at higher fields. This difference originates from the distinct energy-conservation conditions governing the two processes. For MC processes, energy conservation reads $\Delta + \hbar\omega_q  = \varepsilon({\bf q})$  where $\omega_{\bf q}$ is the phonon frequency and $\varepsilon({\bf q})$ is the energy of the final magnon state. At low magnetic fields, both $\Delta$ and $\varepsilon({\bf q})$ increase with the field in a similar manner, such that the participating magnon and phonon states remain nearly unchanged compared to the zero-field case. In contrast, for MNC processes the energy-conservation condition is $\Delta + \varepsilon({\bf q}) = \hbar\omega_q$. In this case, the magnitude of the phonon wave vector $|{\bf q}|$ satisfying the conservation law decreases with increasing magnetic field, leading to a deviation from Gilbert behavior at moderately high fields.

Finally, Fig.~\ref{fig:SI-YIG}(g) shows the relative contributions of different phonon branches to the magnon–phonon damping, including both MC and MNC processes. As expected, the dominant contribution arises from the acoustic phonon branches; however, several optical branches also provide a small but non-negligible contribution.

\subsection{Benchmarking Magnon–Phonon Damping in YIG}

To the best of our knowledge, there are no studies explicitly focused on the Gilbert damping parameter of ${\bf k}=0$ magnons in YIG or in a monolayer of CrSBr. Nevertheless, at least two works~\cite{goodYIG19,goodYIG2014} allow one to extract this parameter for the magnon–phonon contribution to the damping in YIG. In this section, we compare our results with the findings of these studies.

An estimate of the magnon damping parameter can be obtained from Ref.~\cite{goodYIG19}, where magnon–phonon relaxation in YIG was analyzed using a simplified model tuned to reproduce key experimental observations, including the magnon dispersion measured by inelastic neutron scattering and the magnetic Grüneisen parameter. In particular, Fig.~10 of Ref.~\cite{goodYIG19} shows the magnetic-field dependence of the relaxation rate of the ${\bf k}=0$ acoustic magnon at $T=50,\mathrm{K}$. At low magnetic fields, the relaxation rate is linear in the applied field, corresponding to Gilbert-type damping. From this figure, one can estimate
$\gamma^c/B \sim 2.5\times 10^5 {\rm s^{-1}T^{-1}}$ and $\gamma^{nc}/B \sim 4.8\times 10^5 {\rm s^{-1}T^{-1}}$.
where $\gamma^{\mathrm{c}}$ and $\gamma^{\mathrm{nc}}$ denote the relaxation rates due to MC and MNC processes, respectively. Assuming that the magnon gap arises solely from the applied magnetic field -— which is reasonable for an order-of-magnitude estimate -— these values correspond to a Gilbert damping parameter $\alpha \simeq 4.1 \times 10^{-8}$, with the MNC contribution approximately twice as large as that from MC processes. In comparison, our calculations yield $\alpha \simeq 1.3 \times 10^{-8}$ at $T=50,\mathrm{K}$, with nearly equal contributions from MC and MNC processes.

A similar comparison can be made with Ref.~\cite{goodYIG2014}, where Fig.~7 reports the relaxation rates of magnons down to ${\bf k}=0$ in a $6.7,\mu\mathrm{m}$-thick YIG film subjected to an external magnetic field of $0.171,\mathrm{T}$ at $T=300,\mathrm{K}$. Although the absence of a magnetic-field dependence precludes a direct assessment of the Gilbert or non-Gilbert character of the damping, the reported relaxation rates allow one to estimate a damping parameter of $\alpha \simeq 1.6 \times 10^{-7}$. By contrast, our calculations yield $\alpha \simeq 0.8 \times 10^{-7}$ at this temperature.

Overall, we consider the agreement among the three approaches to be reasonable, given the substantial differences in the underlying methodologies and geometries, including simplified models fitted to experimental data versus ab initio–based calculations, as well as finite-thickness YIG films versus an idealized infinite YIG crystal.

\section{Magnon-Magnon Damping mechanism in ${\rm CrI_3}$ and ${\rm CrPS_4}$}

\begin{figure*}
 	\centering
 	\includegraphics[width=4.5in]{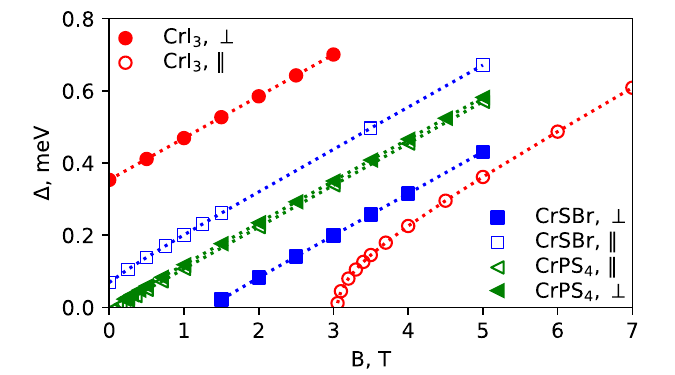}
 	\caption{ \label{fig:SI-BD}  Calculated dependence of magnon gap $\Delta$ on the applied magnetic field $B$. The field is assumed to be in-plane for parallel configurations ($\|$) and out-of-plane  for perpendicular configurations ($\perp$). } 
\end{figure*}

Here we present detailed calculations of the damping in ${\rm CrI_3}$ and ${\rm CrPS_4}$ and compare the results with those obtained for ${\rm CrSBr}$. We use the spin Hamiltonian from~\cite{CrI3model} for ${\rm CrI_3}$ and the Hamiltonian from~\cite{CrPS4model} for ${\rm CrPS_4}$. In both the cases the provided spin Hamiltonians are considered to represent Heisenberg exchange interaction and MCA. The dipole-dipole interaction is added to the model based on the positions of magnetic atoms.

We focus on four-magnon (MC) processes, which provide the dominant contribution to the damping in both YIG and ${\rm CrSBr}$. Both ${\rm CrI_3}$ and ${\rm CrPS_4}$ exhibit easy-axis anisotropy, with the preferred magnetization direction perpendicular to the plane. However, a uniform in-plane magnetized state can be stabilized by applying a sufficiently strong external magnetic field. In ${\rm CrPS_4}$, an external field is also required to stabilize a uniform out-of-plane magnetization, since at $B=0$ dipole–dipole interactions favor a multidomain magnetic configuration. Figure~\ref{fig:SI-BD} shows the calculated magnetic-field dependence of the magnon gap for ${\rm CrI_3}$, ${\rm CrPS_4}$, and ${\rm CrSBr}$ for both magnetization orientations. In all cases, the applied field is chosen to stabilize a uniform magnetic configuration: the field is applied in-plane for parallel ($\parallel$) magnetization and out-of-plane for perpendicular ($\perp$) magnetization. For ${\rm CrSBr}$, the in-plane magnetization corresponds to the crystallographic $b$ direction~\cite{CrSBrStrain}.

\begin{figure*}
 	\centering
 	\includegraphics[width=6.0in]{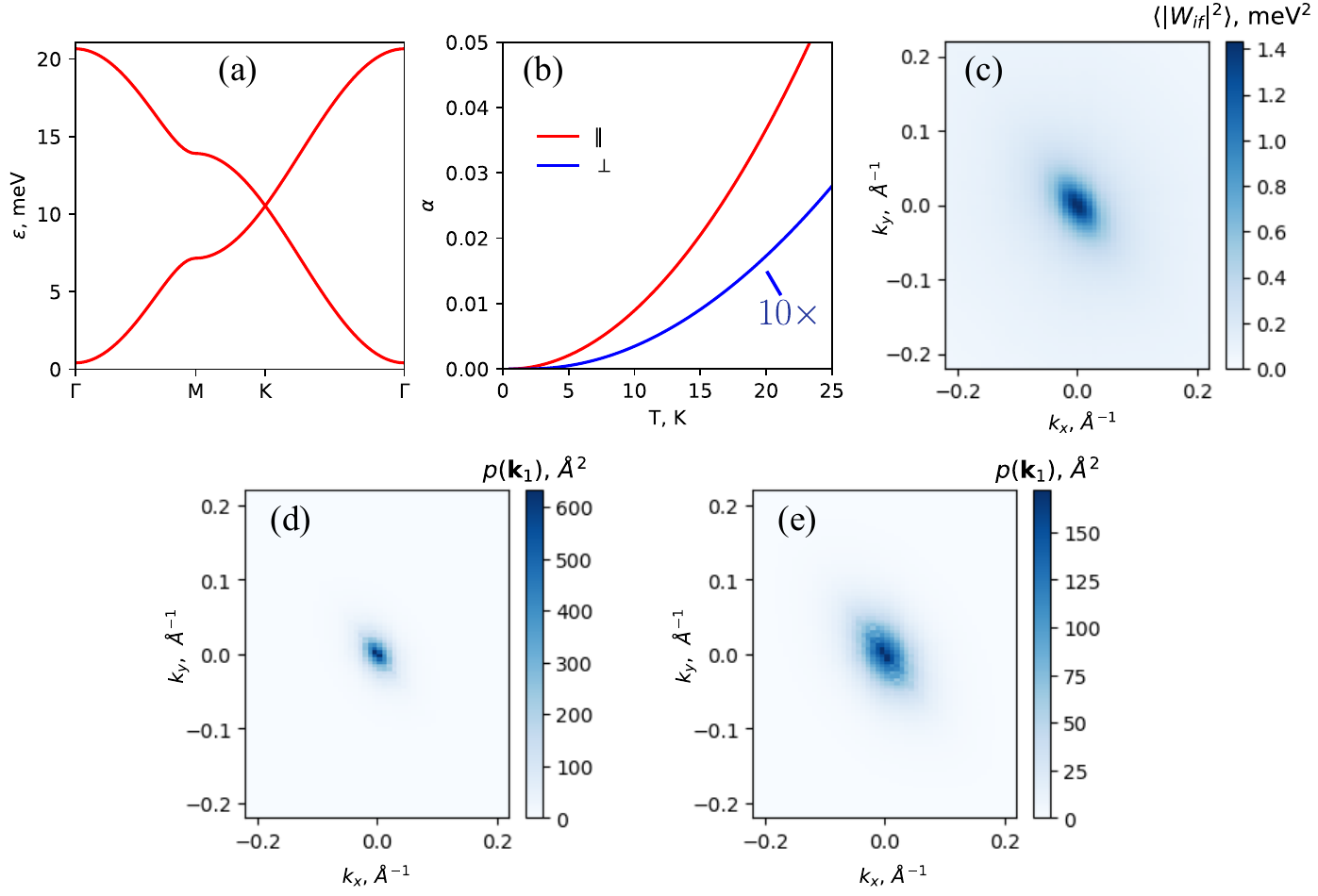}
 	\caption{ \label{fig:SI-CrI3}   Magnon-magnon damping mechanism in ${\rm CrI_3}$. (a) Magnon dispersion derived from the spin Hamiltonian reported in \cite{CrI3model}. (b) Temperature dependence of the damping coefficient $\alpha$ calculated for the external field $B=4\,{\rm T}$ stabilizing either parallel or perpendicular magnetization. (c) The averaged squared matrix elements $|W|^2$ calculated for parallel magnetization in $B = 3.3\,{\rm T}$. (d,e) relative contributions $p({\bf k}_1)$ of different wavevectors ${\bf k}_1$ to the damping in ${\rm CrI_3}$ with parallel magnetization for $B = 3.3\,{\rm T}$ and $B = 4\,{\rm T}$, respectively.} 
\end{figure*}

The magnon dispersion of ${\rm CrI_3}$ calculated using the spin Hamiltonian of Ref.~\cite{CrI3model} is shown in Fig.~\ref{fig:SI-CrI3}(a). It features a sizable magnon gap exceeding $0.4\,{\rm meV}$, reflecting strong out-of-plane anisotropy. While this gap is somewhat reduced when dipole–dipole interactions are included, the uniform ground state remains stable in the absence of an external magnetic field. In contrast, stabilizing a uniform in-plane magnetization requires a strong in-plane field of approximately $3.2\,{\rm T}$. Figure~\ref{fig:SI-CrI3}(b) shows the temperature dependence of the damping coefficient $\alpha$ calculated at $B=4\,{\rm T}$ for both magnetization orientations. In both cases, $\alpha$ follows a $T^2$ scaling. As discussed in the main text, the damping in perpendicularly magnetized ${\rm CrI_3}$ is consistent with the predictions of the toy model. However, for the in-plane configuration the FMR linewidth exhibits a much stronger dependence on the magnon gap than predicted, scaling as $\delta\varepsilon \propto \Delta^{-2.5}$. We attribute this deviation to the contribution of dipole–dipole interactions to the matrix elements $W_{if}$, which acquire a pronounced dependence on the absorbed magnon momentum ${\bf k}_1$.
To clarify this behavior, Fig.~\ref{fig:SI-CrI3}(c) shows the squared matrix elements $|W_{if}|^2$, averaged over the outgoing magnon momenta ${\bf k}_2$ and ${\bf k}_3$, as a function of ${\bf k}_1$ for in-plane magnetization at $B=3.3\,{\rm T}$. The dependence $|W_{if}|^2({\bf k}_1)$ exhibits a pronounced enhancement at small momenta, $|{\bf k}_1| \lesssim 0.04,\text{\AA}^{-1}$. Figures~\ref{fig:SI-CrI3}(d,e) further show the relative contributions $p({\bf k}_1)$ of different incoming momenta to the damping at $B=3.3\,{\rm T}$ and $B=4\,{\rm T}$, respectively. As the magnetic field is reduced, the weight of low-${\bf k}_1$ contributions increases, leading to larger effective values of $|W_{if}|^2$ and, consequently, to a stronger-than-expected enhancement of the damping at small $\Delta$.

The magnon dispersion of ${\rm CrPS_4}$, calculated using the spin Hamiltonian from Ref.~\cite{CrPS4model}, is shown in Fig.~\ref{fig:SI-CrPS4}(a). In this material, a relatively small external magnetic field of approximately $0.1\,{\rm T}$ is required to stabilize a uniform magnetization for both in-plane and out-of-plane configurations.
The temperature dependence of the damping coefficient $\alpha$, calculated at an external field of $4\,{\rm T}$ for both magnetization directions, is presented in Fig.~\ref{fig:SI-CrPS4}(b). In both cases, the damping follows $\alpha \propto T^2$. As discussed in the main text, the dependence of the FMR linewidth on the magnon gap, $\delta\varepsilon(\Delta)$, follows the toy-model prediction $\delta\varepsilon \propto \Delta^{-1}$ only for the in-plane magnetization configuration.

To investigate the origin of this behavior, we plot the momentum-resolved relative contributions $p({\bf k}_1)$ to the damping for perpendicular and parallel magnetization in Figs.~\ref{fig:SI-CrPS4}(c) and (d), respectively. In contrast to the in-plane case and to the behavior expected from the toy model, the contributions for perpendicular magnetization [Fig.~\ref{fig:SI-CrPS4}(c)] exhibit a pronounced minimum near ${\bf k}1=0$. We attribute this feature to a corresponding minimum in the matrix elements $|W{if}|^2$, shown in Fig.~\ref{fig:SI-CrPS4}(e). This minimum originates from destructive interference between the dipole--dipole and magnetocrystalline anisotropy (MCA) contributions to the scattering matrix elements. To confirm this, we calculated the damping in two modified ${\rm CrPS_4}$ models with perpendicular magnetization, in which either the MCA or the dipole--dipole interaction was removed from the Hamiltonian. The resulting FMR linewidths are shown in Fig.~\ref{fig:SI-CrPS4}(g). In both cases, the linewidth recovers the toy-model scaling $\delta\varepsilon \propto \Delta^{-1}$ and increases by almost two orders of magnitude compared to the full model, which includes both MCA and dipole--dipole interactions. These results strongly support the interpretation that the anomalous behavior in the full model arises from destructive interference between these two contributions to the magnon--magnon scattering matrix elements.

\begin{figure*}
 	\centering
 	\includegraphics[width=7.0in]{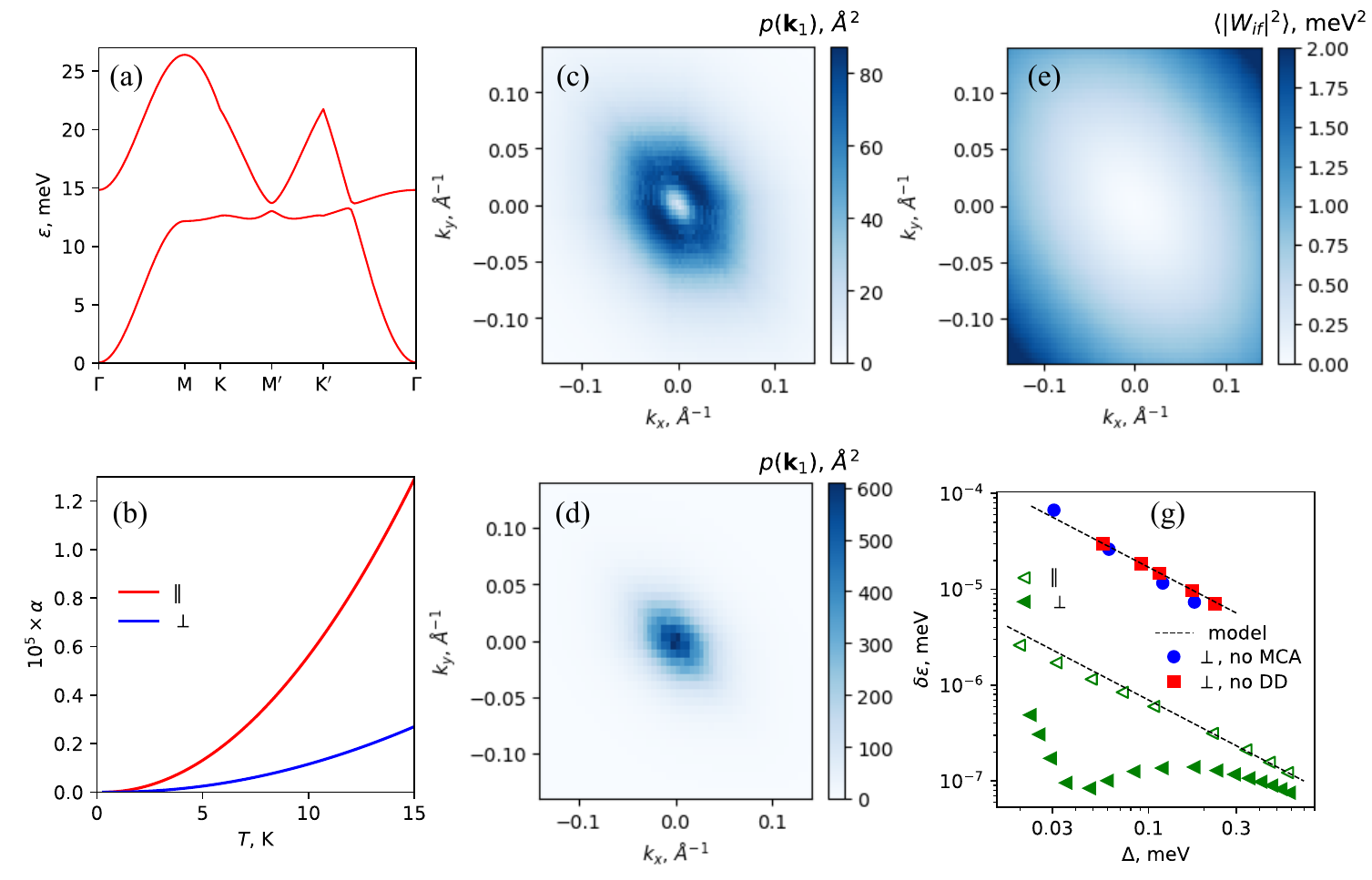}
 	\caption{ \label{fig:SI-CrPS4}   Magnon-magnon damping mechanism in ${\rm CrPS_4}$. (a) The magnon dispersion derived from the spin Hamiltonian published in~\cite{CrPS4model}. (b) The temperature dependence of the damping coefficient $\alpha$ calculated at an external magnetic field of $4\,{\rm T}$ in parallel and perpendicular configurations. (c) The relative contributions of different wavevectors ${\bf k}_1$ in perpendicular configuration calculated at $B=4\,{\rm T}$ and $T=10\,{\rm K}$. (d) The relative contributions of different wavevectors ${\bf k}_1$ in parallel configuration calculated at $B=4\,{\rm T}$ and $T=10\,{\rm K}$. (e) dependence of the averaged squared matrix elements $\langle |W_{if}|^2 \rangle$ on the wavevector ${\bf k}_1$ of the absorbed magnon. (g) FMR linewidth calculated with MCA or dipole-dipole interaction removed from the model compared with the results of the initial model. } 
\end{figure*}

In Fig.~3 of the main text, we compare the magnon--magnon damping calculated for vdW materials with the predictions of the toy model. The original expression [Eq.~(6) of the main text] is formulated in terms of parameters specific to a single-sublattice square-lattice model. To enable comparison with realistic materials, we re-express these parameters in terms of the Curie temperature obtained within molecular-field theory,
\begin{equation}\label{SI-simpleTc}
T_c = \frac{2DS(S+1)}{3}J
\end{equation} 
where $D$ denotes the system dimensionality.

Substituting Eq.~(\ref{SI-simpleTc}) into the toy-model expression yields the following formula for the FMR linewidth:
\begin{equation}\label{SI-mm2Dmod}
\delta\varepsilon = \frac{0.06 T^2 (S+1)^2 \Delta_0^2 }{(T_cS)^2 \Delta}.
\end{equation} 
In Eq.~(\ref{SI-mm2Dmod}), the Curie temperature $T_c$ and the spin $S$ are typically known from experiment, whereas $\Delta_0$ is the only free fitting parameter.

\begin{table}
    \centering
    \begin{tabular}{|c|ccc|}
        \hline
         & $K$, meV $\quad$ & $\Delta_0^{(\perp)}$, meV $\quad$ &  $
         \Delta_0^{(\|)}$, meV \\
         \hline
         ${\rm CrSBr}$ & 0.1 & 0.032 & 0.016\\
         \hline
         ${\rm CrI_3}$ & 0.26 & 0.19 & - \\
         \hline
         ${\rm CrPS_4}$ & 0.006  & 0.0018 & 0.0022 \\
         \hline
    \end{tabular}
    \caption{The fitted values of $\Delta_0$ compared with the energy of magnetic anisotropy $K$. $\Delta_0^{\perp}$ corresponds to a materials with perpendicular magnetization, 
    $\Delta_0^{\|}$ corresponds to parallel magnetization. }
    \label{tab:Delta0}
\end{table}

The fitted values of $\Delta_0$ are listed in Table~\ref{tab:Delta0}. Here, $\Delta_0^{\perp}$ denotes perpendicular magnetization, while $\Delta_0^{\parallel}$ corresponds to in-plane magnetization. For ${\rm CrPS_4}$ in the perpendicular configuration, we use the value of $\Delta_0^{\parallel}$ extracted from high-field calculations, where the results follow Eq.~(\ref{SI-mm2Dmod}). No value of $\Delta_0^{\parallel}$ is reported for ${\rm CrI_3}$, since the corresponding dependence $\delta\varepsilon(\Delta)$ deviates from Eq.~(\ref{SI-mm2Dmod}) across all magnetic fields considered.

We compare the extracted values of $\Delta_0$ with the magnetic anisotropy energy $K$, defined as the absolute difference in total energy between in-plane and out-of-plane magnetization per magnetic atom. This quantity is often available from experiments or \textit{ab initio} calculations without requiring an explicit spin-Hamiltonian derivation. In all cases, $\Delta_0$ is of the same order of magnitude as $K$, with typical values $\Delta_0 \sim 0.3K$, or
\begin{equation}
0.16K < \Delta_0 < 0.7K.
\end{equation}
This observation enables a first-order estimate of the magnon--magnon damping using minimal input data.

\end{widetext}

\end{document}